\documentclass[%
 nofootinbib,
 nobibnotes,
 amsmath,amssymb,
 aps,
]{revtex4-1}

\usepackage{graphicx}
\usepackage{amsmath}
\usepackage{amssymb}
\usepackage{bm}

\begin{document}

\title{Anisotropic Mobility Model for Polymers under Shear \\
and its Linear
Response Functions}

\author{Takashi Uneyama}
 \email{uneyama@scl.kyoto-u.ac.jp}
\affiliation{%
 JST-CREST and
 Institute for Chemical Research, Kyoto University, \\
 Gokasho, Uji, Kyoto 611-0011, Japan
}%
\author{Kazushi Horio}
\author{Hiroshi Watanabe}
\affiliation{%
 Institute for Chemical Research, Kyoto University, \\
 Gokasho, Uji, Kyoto 611-0011, Japan
}%

\date{\today}

\begin{abstract}
We propose a simple dynamic model of polymers under shear with an
 anisotropic mobility tensor.
We calculate the shear viscosity, the rheo-dielectric
response function, and the parallel relaxation modulus under shear flow
 deduced from our model.
We utilize recently developed linear response theories for
nonequilibrium systems to calculate linear response functions.
Our results are qualitatively consistent with experimental results.
We show that our anisotropic mobility model
can reproduce essential dynamical nature of polymers under shear
qualitatively. We compare our model with other models or theories such
 as the convective constraint release model or nonequilibrium linear
 response theories.
\end{abstract}

\maketitle


\section{Introduction}

The linear response theory gives a formula which relates a response
function in equilibrium to an equilibrium time correlation function \cite{Kubo-Toada-Hashitsume-book,Evans-Morris-book,Risken-book}.
In equilibrium, the response of a physical quantity $A$ at time $t$ to a
weak external perturbation at time $t' (\le t)$, 
which is conjugate to $B$, is given as the following form.
\begin{equation}
 \label{linear_response_formula_equilibrium}
 \mathcal{R}_{AB}(t - t')
 = \frac{1}{k_{B} T} \frac{d}{dt'} \langle A(t) B(t') \rangle_{\text{eq}}
\end{equation}
Here $k_{B}$ is the Boltmzann constant, $T$ is the absolute
temperature, and $\langle \dotsb \rangle_{\text{eq}}$ means the equilibrium
statistical average.
Eq \eqref{linear_response_formula_equilibrium} holds for a wide range of
systems \cite{Kubo-Toada-Hashitsume-book,Evans-Morris-book,Risken-book}
as long as the system is in equilibrium.
(In the followings, we call the response formula
\eqref{linear_response_formula_equilibrium} as the Green-Kubo type formula.)
From the view point of experiments, eq
\eqref{linear_response_formula_equilibrium} can be utilized to obtain
the correlation function from the response function.
This enables us to extract
information of microscopic and/or mesoscopic dynamics of molecules
from macroscopic responses.

For polymeric systems, the viscoelastic response functions (especially the
shear relaxation modulus or the storage and loss moduli) or the
dielectric response function are useful.
For example, the viscoelastic response functions can be related to the autocorrelation function
of the microscopic stress tensor, and the stress tensor can be related to
bond vectors \cite{Doi-Edwards-book}.
The dielectric response functions of polymers with type A dipoles
(polymer chains which have electric dipoles along the chain backbones)
can be related to the autocorrelation functions of end-to-end vectors.
The combination of the viscoelastic and dielectric measurements provides
various information about dynamics of polymer chains
\cite{Watanabe-1999,Matsumiya-Watanabe-Osaki-2000,Watanabe-Matsumiya-Osaki-2000,Matsumiya-Watanabe-2001,Watanabe-Matsumiya-Inoue-2002}.

Even out of equilibrium, it is possible to measure linear response
functions. Then we expect that the measured response functions can be
related to the correlation functions, or at least they reflect the information about
the dynamics of polymer chains. Actually, several linear response
measurements under steady shear, such as mechanical responses
\cite{Osaki-Tamura-Kurata-Kotaka-1965,Costello-1977,Vermant-Walker-Moldenaers-Mewis-1998,Isayev-Wong-1988,Wong-Isayev-1989,Somma-Valentino-Titomanlio-Ianniruberto-2007,Boukany-Wang-2009,Li-Wang-2010}
or dielectric responses \cite{Matsumiya-Watanabe-Inoue-Osaki-Yao-1998,Watanabe-Sato-Matsumiya-Inoue-Osaki-1999,Watanabe-Ishida-Matsumiya-2002,Watanabe-Matsumiya-Inoue-2003,Watanabe-Matsumiya-Inoue-2005}
have been reported. However, unlike the equilibrium cases, it is not clear how we
can analyze and interpret the nonequilibrium linear response functions.
In several works, the Green-Kubo type relation
\eqref{linear_response_formula_equilibrium} is utilized to
analyze obtained experimental data, without justifications.
But from the view point of nonequilibrium statistical physics, generally eq
\eqref{linear_response_formula_equilibrium} does not hold
except for some special or limited cases.

Fortunately, the Green-Kubo type relation approximately holds for
the dielectric response of polymers in the shear gradient direction under shear 
(the rheo-dielectric response)
\cite{Uneyama-Masubuchi-Horio-Matsumiya-Watanabe-Pathak-Roland-2009},
and thus we can obtain the correlation function of the end-to-end
vectors under shear.
In this work we therefore concentrate on polymers with type A dipoles.
The rheo-dielectric responses for polymers with type A dipoles
has been systematically studied and analyzed for various systems and
shear rates
\cite{Matsumiya-Watanabe-Inoue-Osaki-Yao-1998,Watanabe-Ishida-Matsumiya-2002,Watanabe-Matsumiya-Inoue-2003,Watanabe-Matsumiya-Inoue-2005},
The rheo-dielectric functions of linear polymers are reported to be insensitive to shear,
even if the shear rate is large and the shear thinning is observed.
This implies that the dynamics of polymer chains in the shear gradient
direction is not
affected largely by shear flow even under fast shear.
So far, why and how such insensitivity occurs is not fully understood.
Although attempts have been done by coarse-grained molecular
simulations
\cite{Masubuchi-Watanabe-Ianniruberto-Greco-Marrucci-2004,Uneyama-Masubuchi-Horio-Matsumiya-Watanabe-Pathak-Roland-2009},
simulations have failed to reproduce experimentally
obtained rheo-dielectric behavior.


In the field of the constitutive equation models, many different
approaches to nonequilibrium systems have been utilized, to reproduce
nonlinear viscoelasticity well.
Among them, the anisotropic
mobility type models
\cite{Curtiss-Bird-1980,Curtiss-Bird-1980a,Giesekus-1982,Arsac-Carrot-Guillet-Revenu-1994,Stephanou-Baig-Mavrantzas-2009,Beris-Edwards-book}
are particularly notable.
In the anisotropic friction model, the friction
coefficient (or almost equivalently, the relaxation time) of the model
is expressed as a tensor quantity instead of a scalar quantity. This
allows us to reproduce variety of models which can
reproduce complex viscoelastic behavior with relatively simple
constitutive equations.

Motivated by the anisotropic mobility type constitutive equation models,
in this work we aim to propose a Langevin equation model with an anisotropic
mobility tensor for dynamics of polymers under shear.
Although the anisotropic mobility tensors are not
utilized widely in the field of the nonequilibrium statistical
mechanics, they can potentially model the dynamics under fast shear.
A Langevin equation model with an anisotropic mobility tensor model was
already proposed for a rather simple system
\cite{McPhie-Daivis-Snook-Ennis-Evans-2001} based on the projection
operator method and molecular dynamics simulation data.
We construct an anisotropic mobility model which is
consistent with some molecular dynamics simulation results, in a similar
way.
We limit ourselves to an analytically solvable toy model to examine the
characteristic properties of the anisotropic mobility model explicitly and simply.
We explicitly calculate several linear response functions such as the
rheo-dielectric response function and the moduli in the shear flow
direction (parallel moduli) for our model.
To calculate the linear response functions, we utilize recently developed
linear response theories for nonequilibrium Langevin systems
\cite{Baiesi-Maes-Wynants-2009,Baiesi-Maes-Wynants-2009a,Seifert-Speck-2010}.
Finally we discuss the properties of our model and its linear
response functions under shear from several aspects. We compare our
model with several pieces of previous work, and show the differences and
similarities between our model and the other models.

\section{Model}

In this section, we propose a simple and solvable model for dynamics of
polymers under shear. We consider weakly entangled polymer melts or
solutions, for which rheo-dielectric experiments have been carried out \cite{Matsumiya-Watanabe-Inoue-Osaki-Yao-1998,Watanabe-Ishida-Matsumiya-2002,Watanabe-Matsumiya-Inoue-2003,Watanabe-Matsumiya-Inoue-2005}.
There are several different
approaches to model the dynamics of polymers
\cite{Doi-Edwards-book,Watanabe-1999,McLeish-2002,Kroger-2004}. Thus at first, we
should choose an appropriate model from candidates. We are
interested in the qualitative and essential feature of the polymer dynamics
under shear.
We require the model to be simple and analytically solvable, so that we
avoid unnecessary complexities and confusions, and make the model
properties clear. From these requirements, we limit ourselves to the
dynamics of the end-to-end vector of a single polymer chain $\bm{R}$,
in the weakly entangled system. This
reduces the degrees of freedom drastically. (This
approximation gives so-called the dumbbell model \cite{Kroger-2004}.)
We also limit ourselves to the dynamics in the long time limit, where is
no memory effect for the end-to-end vector dynamics. Then we can describe
the dynamic equation for $\bm{R}$ in a closed, Markovian form.
In equilibrium, this equation can be described as
\begin{equation}
 \label{langevin_equation_equilibrium}
 \frac{d\bm{R}(t)}{dt} =
  - \frac{1}{\zeta_{0}} \frac{\partial
  \mathcal{F}(\bm{R}(t))}{\partial \bm{R}(t)}
  + \bm{\xi}_{0}(t)
\end{equation}
where $\zeta_{0}$ is the friction coefficient which the end-to-end
vector feels, $\mathcal{F}(\bm{R})$ is the free energy, and
$\bm{\xi}_{0}(t)$ is the Gaussian noise. For a Gaussian chain, the
free energy can be expressed in the following simple linear elasticity form.
\begin{equation}
 \label{free_energy_linear_elasticity}
 \mathcal{F}(\bm{R}) = \frac{3 k_{B} T}{2 \bar{R}^{2}} \bm{R}^{2}
\end{equation}
$\bar{R}^{2}$ is the equilibrium mean square average end-to-end
distance. The fluctuation-dissipation relation of the second kind \cite{Kubo-Toada-Hashitsume-book}
requires for the thermal noise $\bm{\xi}_{0}(t)$ to satisfy the following relations.
\begin{align}
 & \langle \bm{\xi}_{0}(t) \rangle = 0 \\
 & \langle \bm{\xi}_{0}(t) \bm{\xi}_{0}(t') \rangle = \frac{2 k_{B} T}{\zeta_{0}}
 \delta(t - t') \bm{1}
\end{align}
Here $\langle \dotsb \rangle$ denotes the statistical average and
$\bm{1}$ is the unit tensor.

\begin{figure}[h!]
\includegraphics{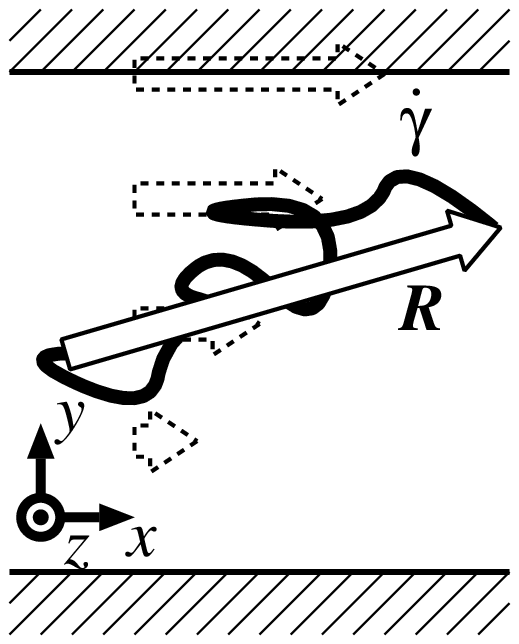}
\caption{\label{schematic_image_polymer_under_shear}}
\end{figure}

We consider a system under simple shear.
The velocity gradient tensor $\bm{\kappa}$
is given as follows.
\begin{equation}
 \label{velocity_gradient_tensor_simple_shear}
 \kappa_{\alpha \beta} = 
  \begin{cases}
   \dot{\gamma} & (\alpha = x, \beta = y) \\
   0 & (\text{otherwise})
  \end{cases}
\end{equation}
where $\dot{\gamma}$ is the shear rate.
We assume that $\dot{\gamma}$ is smaller than $\tau_{R}^{-1}$
($\tau_{R}$ is the Rouse time of the polymer chain) and the polymer
chain is not so highly stretched.
A schematic image of the system
is shown in FIG. \ref{schematic_image_polymer_under_shear}.
Since the shear flow cannot be expressed in relation to a conservative potential force, the
system is nonequilibrium under simple shear. As a simple extension of
the equilibrium Langevin equation \eqref{langevin_equation_equilibrium},
one may consider the following Langevin equation.
\begin{equation}
 \label{langevin_equation_equilibrium_extended_to_under_shear}
 \frac{d\bm{R}(t)}{dt} =
  - \frac{1}{\zeta_{0}} \frac{\partial
  \mathcal{F}(\bm{R}(t))}{\partial \bm{R}(t)}
  + \bm{\kappa} \cdot \bm{R}(t)
  + \bm{\xi}_{0}(t)
\end{equation}
In the Langevin equation \eqref{langevin_equation_equilibrium_extended_to_under_shear}, the
mobility, which is defined as the inverse of the friction coefficient,
is scalar and thus isotropic. This is because we simply used the
mobility (the friction tensor) in equilibrium. The system is
isotropic in equilibrium, and the mobility should also be isotropic from the symmetry.
However, if the system is not in or
near equilibrium, it is generally not isotropic. This means
that, in principle, the mobility could be an anisotropic, tensor
quantity \cite{McPhie-Daivis-Snook-Ennis-Evans-2001}.
Actually, in some constitutive equation models
\cite{Giesekus-1982,Beris-Edwards-book}, the mobility tensors are
designed to be anisotropic under flow.
Thus we consider that eq
\eqref{langevin_equation_equilibrium_extended_to_under_shear} is an
oversimplified dynamic equation model, which may lead
physically incorrect results in particular under fast flow.

When the mobility tensor becomes anisotropic, we expect that
some transport coefficient tensors also become anisotropic \cite{Sarman-Evans-Cummings-1991,Evans-1991,Evans-Baranyai-Sarman-1992}.
Anisotropic diffusion coefficient tensors are actually observed in
nonequilibrium molecular dynamics (NEMD) simulations based on the
SLLOD model \cite{Sarman-Evans-Baranyai-1992,Hunt-Todd-2009}.
The diffusion tensor of Lennard-Jones (LJ) particles under shear is
first reported by Sarman, Evans, and Baranyai \cite{Sarman-Evans-Baranyai-1992}. They studied the
diffusion tensor of an LJ particle under shear, at various 
rates. Their simulation result clearly shows that the diffusion tensor becomes anisotropic and
it depends on the shear rate.
Quite recently, Hunt and Todd \cite{Hunt-Todd-2009} performed NEMD simulations for relatively short
polymer melts (the Kremer-Grest chains \cite{Kremer-Grest-1990}) and
measured the diffusion tensors under shear. Their
result is qualitatively similar to the case of the LJ system. Namely,
the diffusion tensor of the center of mass of a polymer chain becomes
anisotropic and depends on the shear rate. Besides, the dependence of
the diffusion tensor on the shear rate becomes strong as the
polymerization index (number of beads in a chain) increases.
Although their simulations are limited for rather short chains (unentangled
chains), we expect that this trend will be qualitatively the same or
even enhanced for well entangled chains.

Although there are several possible interpretations for these NEMD results \cite{Baranyai-2000,McPhie-Daivis-Snook-Ennis-Evans-2001},
in this work we interpret that the anisotropic diffusion is caused by
anisotropic mobilities.
Namely, if we model the dynamics of a polymer chain under shear by a
Langevin equation, we
should employ an anisotropic mobility tensor (or an
anisotropic friction tensor).
McPhie et al \cite{McPhie-Daivis-Snook-Ennis-Evans-2001} proposed a
Langevin equation with an anisotropic friction tensor to describe the
coarse-grained motion of an LJ particle under shear.
Although they proposed an underdamped Langevin equation,
an overdamped Langevin equation (like eq
\eqref{langevin_equation_equilibrium}) seems to be more suitable for the end-to-end
vector of a polymer.
Then we express the Langevin equation under shear as follows.
\begin{equation}
 \label{langevin_equation_under_shear}
 \frac{d\bm{R}(t)}{dt} =
  - \bm{\Lambda}(\dot{\gamma}) \cdot \frac{\partial
  \mathcal{F}(\bm{R})}{\partial \bm{R}}
  + \bm{\kappa} \cdot \bm{R} + \bm{\xi}(\dot{\gamma};t)
\end{equation}
where $\bm{\Lambda}(\dot{\gamma})$ is the mobility tensor which depends on the
shear rate $\dot{\gamma}$ and $\bm{\xi}(\dot{\gamma},t)$ is the Gaussian
noise. We assume that the fluctuation-dissipation type relation between
$\bm{\xi}(\dot{\gamma},t)$ and $\bm{\Lambda}(\dot{\gamma})$ is satisfied.
\begin{align}
 & \langle \bm{\xi}(\dot{\gamma};t) \rangle = 0 \\
 & \label{fluctuation_dissipation_relation_2nd_moment}
 \langle \bm{\xi}(\dot{\gamma};t) \bm{\xi}(\dot{\gamma};t') \rangle = 2 k_{B} T
 \bm{\Lambda}(\dot{\gamma}) \delta(t - t')
\end{align}
Eq \eqref{fluctuation_dissipation_relation_2nd_moment} requires the
mobility tensor $\bm{\Lambda}(\dot{\gamma})$ to be symmetric under the
shear field given by eq \eqref{velocity_gradient_tensor_simple_shear}.
Quite recently, Ilg and Kr\"{o}ger \cite{Ilg-2010,Ilg-Kroger-2011}
proposed a constitutive equation model with anisotropic mobility
(friction coefficient) tensor for relatively short polymer chains,
based on the NEMD results.
Although their model is not equivalent to ours, it is qualitatively
similar.

From NEMD simulation results and properties of a Langevin type equation,
we consider the mobility tensor
$\bm{\Lambda}(\dot{\gamma})$ under steady shear should have the
following properties.
\begin{itemize}
 \item The mobility tensor should be positive definite (all of its
       eigenvalues should be positive).
 \item The eigenvalues of
       $\bm{\Lambda}(\dot{\gamma})$ are unchanged under the transform
       $\dot{\gamma} \to - \dot{\gamma}$. That is, each eigenvalue is an even
       function of $\dot{\gamma}$.
 \item The $xx$-component ($\Lambda_{xx}$) decreases as $\dot{\gamma}$
       increases, while other diagonal components ($\Lambda_{yy}$ and
       $\Lambda_{zz}$) are not so sensitive to $\dot{\gamma}$.
 \item The $xy$-component ($\Lambda_{xy}$) is non-zero but its value is
       smaller than diagonal elements. Thus it may be simply neglected
       ($\Lambda_{xy} \approx 0$).
 \item From the symmetry, $\Lambda_{xz} = \Lambda_{yz} = 0$.
 \item The mobility tensor should reduce to the isotropic tensor at
       equilibrium ($\dot{\gamma} = 0$).
\end{itemize}
In this work, we employ the following
simple form which satisfies the above properties. (We discuss about
the other possible forms later.)
\begin{equation}
 \label{mobility_tensor_model}
  \bm{\Lambda}(\dot{\gamma}) = \frac{1}{\zeta_{0}}
  \begin{bmatrix}
   \tilde{\lambda}(\dot{\gamma}) & 0 & 0 \\
   0 & 1 & 0 \\
   0 & 0 & 1
  \end{bmatrix}
\end{equation}
Here, $\zeta_{0}$ is the friction coefficient at equilibrium and
$\tilde{\lambda}(\dot{\gamma})$ is a function of
$\dot{\gamma}$. $\tilde{\lambda}(\dot{\gamma})$ is an even function of
$\dot{\gamma}$, and it monotonically increases as $\dot{\gamma}^{2}$
increases. Further, since it should recover the equilibrium form in
the absence of shear flow, $\tilde{\lambda}(\dot{\gamma})$ reduces to the
equilibrium form, $\tilde{\lambda}(0) = 1$, at the limit of $\dot{\gamma} \to 0$.
For example, we can employ the following simple form for
$\tilde{\lambda}(\dot{\gamma})$.
\begin{equation}
 \label{dimensionless_mobility_factor_power_law}
 \tilde{\lambda}(\dot{\gamma}) = \left[ 1 + (\tau_{c}
 \dot{\gamma})^{2} \right]^{\alpha / 2}
\end{equation}
$\alpha$ is an exponent which satisfies $0 \le \alpha \le 1$, and
$\tau_{c}$ is a characteristic crossover time. Roughly speaking,
the mobility tensor is isotropic for $|\tau_{c} \dot{\gamma}| \ll 1$ and
anisotropic for $|\tau_{c} \dot{\gamma}| \gg 1$. As we show in the next
section, eq \eqref{dimensionless_mobility_factor_power_law} gives the
power law type behavior of the shear viscosity. We call eq
\eqref{dimensionless_mobility_factor_power_law} as the power law type
model.
We can also employ the following form.
\begin{equation}
 \label{dimensionless_mobility_factor_second_newtonian}
 \tilde{\lambda}(\dot{\gamma}) = 1 + \frac{\zeta_{0} -
 \zeta_{\infty}}{\zeta_{\infty}} \frac{( \tau_{c} \dot{\gamma})^{2}}{1 +
 (\tau_{c} \dot{\gamma})^{2}}
\end{equation}
$\zeta_{\infty}$ corresponds to the
effective friction coefficient in the $x$-direction for $
|\tau_{c} \dot{\gamma}| \gg 1$. Eq \eqref{dimensionless_mobility_factor_second_newtonian} may be preferred if the system exhibits the
second Newtonian region for the shear viscosity.
Eqs \eqref{dimensionless_mobility_factor_power_law} and
\eqref{dimensionless_mobility_factor_second_newtonian} are just possible
candidates, and there are many other possible forms for
$\tilde{\lambda}(\dot{\gamma})$.
It is worth noting that our anisotropic mobility tensor is qualitatively
similar to the model by Ilg and Kr\"{o}ger
\cite{Ilg-2010,Ilg-Kroger-2011} (in their model, the dynamics of the
$x$-direction is also accelerated under shear).
Before we proceed, we should notice that the anisotropic mobility tensor
\eqref{mobility_tensor_model} is designed for the simple shear flow, and
it is not expected to be applicable for other flows such as elongational flows.
In the following analysis, we consider the response around the
steady state under the simple shear (expressed by eq
\eqref{velocity_gradient_tensor_simple_shear}) and
thus eq \eqref{mobility_tensor_model} is sufficient for our purpose.

Since the Langevin equation \eqref{langevin_equation_under_shear} is
linear in $\bm{R}$, we can analytically integrate it and obtain
explicit expression for several physical quantities. Thus we can analyze
linear responses such as the rheo-dielectric response of our model explicitly.
By substituting eq \eqref{mobility_tensor_model}
into eq \eqref{langevin_equation_under_shear}, we have the following
equations for $R_{x}, R_{y},$ and $R_{z}$.
\begin{align}
 & \label{langevin_equation_under_shear_x}
 \frac{dR_{x}(t)}{dt} =
  - \frac{\tilde{\lambda}(\dot{\gamma})}{\tau_{0}} R_{x}(t)
  + \dot{\gamma} R_{y}(t)  + \xi_{x}(\dot{\gamma},t) \\
 & \label{langevin_equation_under_shear_y}
 \frac{dR_{y}(t)}{dt} =
  - \frac{1}{\tau_{0}} R_{y}(t) + \xi_{y}(\dot{\gamma},t) \\ 
 & \label{langevin_equation_under_shear_z}
 \frac{dR_{z}(t)}{dt} =
  - \frac{1}{\tau_{0}} R_{z}(t)
  + \xi_{z}(\dot{\gamma},t)
\end{align}
where we defined the characteristic time $\tau_{0}$.
\begin{equation}
 \tau_{0} \equiv \frac{\zeta_{0} \bar{R}^{2}}{3 k_{B} T}
\end{equation}
The probability distribution defined via the following equation is useful for
some calculations.
\begin{equation}
 \label{time_dependent_probability_distribution}
 P(\bm{r},t) \equiv \langle \delta(\bm{r} - \bm{R}(t)) \rangle
\end{equation}
The probability density
\eqref{time_dependent_probability_distribution} follows the
Fokker-Planck equation, which describes the time evolution due to the Langevin equation
\eqref{langevin_equation_under_shear}.
\begin{equation}
 \label{fokker_planck_equation_under_shear}
  \frac{\partial P(\bm{r},t)}{\partial t} =
  \frac{\partial}{\partial \bm{r}} \cdot \bm{\Lambda}(\dot{\gamma}) \cdot
  \left[ \frac{\partial \mathcal{F}(\bm{r})}{\partial \bm{r}}
   P(\bm{r},t) + k_{B} T \frac{\partial P(\bm{r},t)}{\partial \bm{r}} \right]
  - \frac{\partial}{\partial \bm{r}} \cdot
  \left[ \bm{\kappa} \cdot \bm{r} P(\bm{r},t) \right]
\end{equation}
The steady state solution of the Fokker-Planck equation
\eqref{fokker_planck_equation_under_shear}, $P_{\text{ss}}(\bm{r})$
satisfies
\begin{equation}
 \label{steady_state_probability_distribution_condition}
  0 =
  \frac{\partial}{\partial \bm{r}} \cdot \bm{\Lambda}(\dot{\gamma}) \cdot
  \left[ \frac{\partial \mathcal{F}(\bm{r})}{\partial \bm{r}}
   P_{\text{ss}}(\bm{r}) + k_{B} T \frac{\partial P_{\text{ss}}(\bm{r})}{\partial \bm{r}} \right]
  - \frac{\partial}{\partial \bm{r}} \cdot
  \left[ \bm{\kappa} \cdot \bm{r} P_{\text{ss}}(\bm{r}) \right]
\end{equation}
For a linear Fokker-Planck equation, the steady state probability
distribution is known to be a
Gaussian \cite{Risken-book}. Thus we can describe the explicit form of
$P_{\text{ss}}(\bm{r})$ simply as follows.
\begin{equation}
 \label{steady_state_probability_distribution}
 P_{\text{ss}}(\bm{r}) =
 \frac{1}{\sqrt{\det [2 \pi \bm{C}(\dot{\gamma})]}}
  \exp\left[ - \frac{1}{2} \bm{r} \cdot \bm{C}^{-1}(\dot{\gamma}) \cdot \bm{r} \right]
\end{equation}
Here $\bm{C}(\dot{\gamma})$ is the covariance matrix.
\begin{equation}
 \label{steady_state_covariance_matrix}
 \bm{C}(\dot{\gamma}) \equiv
  \frac{\bar{R}^{2}}{3}
  \begin{bmatrix}
   \displaystyle 1
  + \frac{(\tau_{0} \dot{\gamma})^{2}}{\tilde{\lambda} (1 + \tilde{\lambda})} &
  \displaystyle \frac{\tau_{0} \dot{\gamma}}{1 + \tilde{\lambda}} & 0 \\
  \displaystyle \frac{\tau_{0} \dot{\gamma}}{1 + \tilde{\lambda}} & 1 & 0 \\
  0 & 0 & 1
  \end{bmatrix}
\end{equation}
For any physical quantity determined by $\bm{R}$, the statistical average of a function of $\bm{R}$ in the steady state
can be evaluated easily with the aid of the steady state probability
distribution \eqref{steady_state_probability_distribution}.
Also, we can construct the constitutive equation of the anisotropic mobility
model from the Fokker-Planck equation
\eqref{fokker_planck_equation_under_shear}.
We show the constitutive equation and compare it with some conventional
constitutive equation
models in Appendix \ref{constitutive_equation_for_anisotropic_mobility_model}.

\section{Results}

\subsection{Shear Viscosity}

In our model, the stress tensor is expressed as follows.
\begin{equation}
 \label{stress_tensor_definition}
 \bm{\sigma}(\bm{R})
  = \frac{3 \nu_{0} k_{B} T}{\bar{R}^{2}} \bm{R} \bm{R}
\end{equation}
where $\nu_{0}$ is the number density of polymer chains.
The steady state statistical average of eq
\eqref{stress_tensor_definition} can be evaluated easily by using eqs
\eqref{steady_state_probability_distribution} and \eqref{steady_state_covariance_matrix}.
The shear viscosity at the steady state is expressed as
\begin{equation}
 \eta(\dot{\gamma}) \equiv \frac{\langle \sigma_{xy} \rangle_{\text{ss}}}{\dot{\gamma}}
\end{equation}
where $\langle \dotsb \rangle_{\text{ss}}$ denotes the statistical
average in the steady state (under shear).
Finally we have the following expression for the steady state shear
viscosity $\eta(\dot{\gamma})$.
\begin{equation}
 \label{steady_state_viscosity}
 \eta(\dot{\gamma})
   = \frac{3 \nu_{0} k_{B} T}{\bar{R}^{2} \dot{\gamma}}
   C_{xx}(\dot{\gamma})
   = \eta_{0} \frac{2}{1 + \tilde{\lambda}}
\end{equation}
Here $\eta_{0}$ is the Newtonian viscosity in the
limit of $\dot{\gamma} \to 0$, $\eta_{0} \equiv \nu_{0}
k_{B} T \tau_{0} / 2$.
Eq \eqref{steady_state_viscosity} is a monotonically decreasing function
of $\dot{\gamma}^{2}$. Thus we find that our model can reproduce the shear
thinning behavior.

Here we note that the shear thinning mechanism of
our model is somehow similar to ones of the Giesekus model \cite{Giesekus-1982}
or the Johnson-Segalman (JS)
model \cite{Johnson-Segalman-1977,Wu-Schummer-1990}.
Both the Giesekus model and the JS model are kind of linear
dumbbell model which describes the dynamic behavior of viscoelastic
fluids. (We briefly compare our
model with the Giesekeus model or the JS model in Appendix
\ref{constitutive_equation_for_anisotropic_mobility_model}.)
In the Giesekus model, the anisotropic mobility comes from the
hydrodynamic interaction from surrounding polymer chains.
The anisotropic mobility tensor depends on the average conformation,
which leads the effective acceleration of the relaxation and the shear
thinning behavior.
(The hydrodynamic interaction for a linear dumbbell model and the
resulting shear thinning behavior have been
extensively studied \cite{Zylka-Ottinger-1989,Wedgewood-1989,Prabhakar-Prakash-2002}.)
On the other hand, the characteristic feature of the JS model is the
slippage effect, which allows a dumbbell to locally slip and effectively
reduces shear deformation.
In both models, a dumbbell apparently relaxes faster as we
increase the shear rate, independent of their detail mechanisms.
Our model behaves in a similar way.
However, in our model, we did not explicitly consider the origin of the anisotropic
mobility tensor \eqref{mobility_tensor_model}.
(In our model, the mobility tensor is designed based on the MD results,
whereas in the conventional models it is usually designed from specific
kinetic interactions.)
We also note that nonlinear elasticity models (such as the finite
extensibility nonlinear elasticity models \cite{Kroger-2004}) can reproduce similar shear
thinning behavior. It is difficult to identify
the molecular level mechanism of shear thinning behavior only from the
shear viscosity data.



\begin{figure}[h!]
\includegraphics{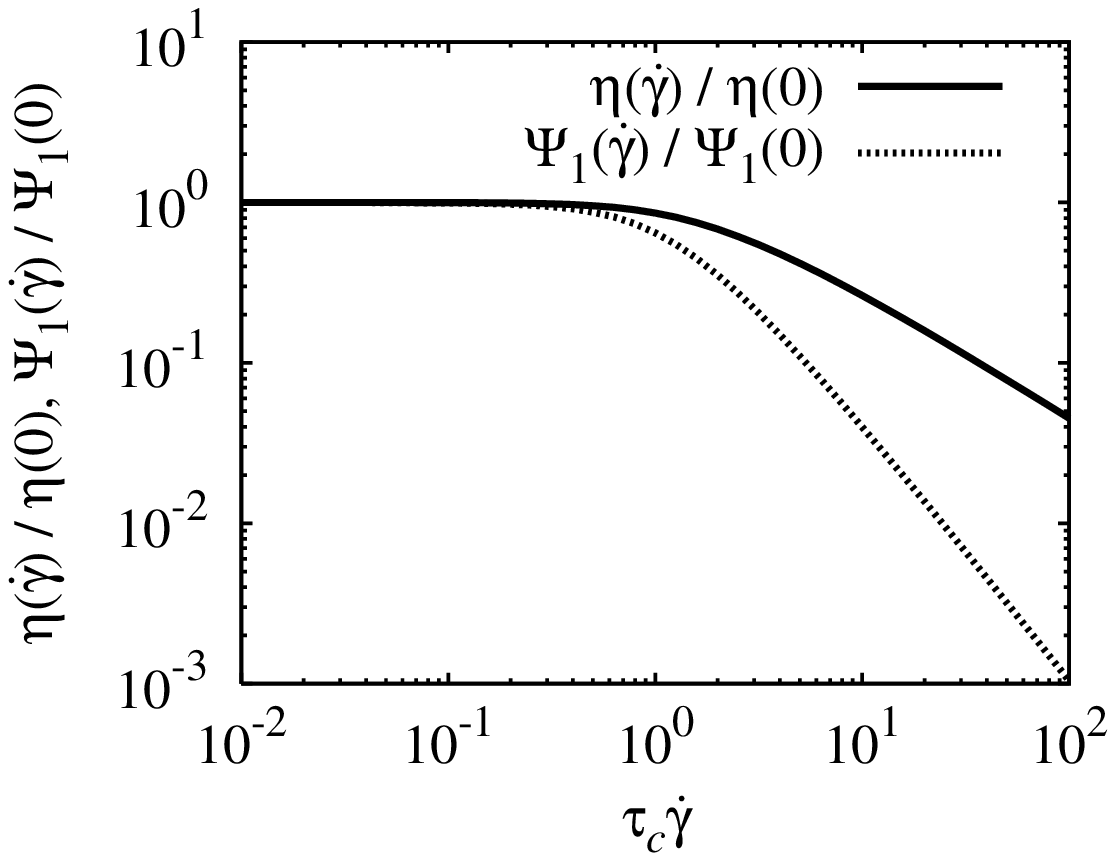}
\caption{\label{eta_and_psi1_power_law}}
\end{figure}

The first normal stress difference coefficient can be calculated in a similar way.
\begin{equation}
 \Psi_{1}(\dot{\gamma})
  \equiv \frac{\langle \sigma_{xx} - \sigma_{yy}
  \rangle_{\text{ss}}}{\dot{\gamma}^{2}}
  = \frac{\nu_{0} k_{B} T \tau_{0}^{2}}{2}
  \frac{2}{\tilde{\lambda} (1 +
  \tilde{\lambda})}
\end{equation}
This also exhibits the thinning behavior. However, the dependences on the
shear rate of $\eta(\dot{\gamma})$ and $\Psi_{1}(\dot{\gamma})$ are different.
If we employ the power law type model for $\tilde{\lambda}$ (eq
\eqref{dimensionless_mobility_factor_power_law}), at the high shear rate
region we have
\begin{align}
 & \eta(\dot{\gamma}) \propto \dot{\gamma}^{- \alpha} \qquad (\dot{\gamma}
 \to \infty) \\
 & \Psi_{1}(\dot{\gamma}) \propto \dot{\gamma}^{- 2 \alpha} \qquad
 (\dot{\gamma} \to \infty)
\end{align}
The shear viscosity and the first normal stress difference coefficient
for the power law type model (with $\alpha = 9 / 11$ \cite{Graessley-1967}) are shown in
FIG. \ref{eta_and_psi1_power_law}.

\subsection{Rheo-Dielectric Response Function}

In dielectric measurements, we impose the time-dependent electric field in
the $y$-direction (shear gradient direction), $E_{y}(t)$, and we measure the $y$-component of the electric flux density
density. For a polymer chain which has a type-A dipole,
the electric flux density (strictly speaking, the $y$-component
of the electric flux density) $D_{y}$ can be expressed as follows \cite{Kubo-Toada-Hashitsume-book}.
\begin{equation}
 D_{y}(\bm{R}) = \varepsilon_{\infty} E_{y} + 4 \pi \nu_{0} \tilde{\mu} R_{y}
\end{equation}
where $\varepsilon_{\infty}$ is the effective dielectric constant due to
the fast dynamics which is not resolved in our model,
and $\tilde{\mu}$ is the effective dipole intensity per unit backbone
length of a polymer chain.
At equilibrium, the Green-Kubo type formula (which is also referred as
the Cole formula) gives
\begin{equation}
 \label{dielectric_response_function_green_kubo}
 \varphi(t - t') = \frac{1}{k_{B} T} \bigg\langle [4 \pi \nu_{0} \tilde{\mu}
  R_{y}(t)] \frac{d}{dt'}[ \tilde{\mu} R_{y}(t')] \bigg\rangle_{\text{eq}}
  = \frac{4 \pi \nu_{0} \tilde{\mu}^{2}}{k_{B} T} \bigg\langle
  R_{y}(t) \frac{dR_{y}(t')}{dt'} \bigg\rangle_{\text{eq}}
\end{equation}
Here, for simplicity we assumed that the correction factor for the
internal electric field is unity.
In equilibrium, our model (which reduces to eq \eqref{langevin_equation_equilibrium}) gives a single Debye type dielectric relaxation function. The
result is
\begin{equation}
 \label{dielectric_response_function_equilibrium}
 \varphi(t)
  = \Delta\varepsilon_{0} \frac{1}{\tau_{0}}
  e^{- t / \tau_{0}}
\end{equation}
where we defined the dielectric intensity $\Delta\varepsilon_{0}$ as
$\Delta\varepsilon_{0} \equiv 4 \pi \nu_{0} \tilde{\mu}^{2} \bar{R}^{2} / 3 k_{B} T$.

Under shear, generally we cannot use the Green-Kubo type response formula
\eqref{dielectric_response_function_green_kubo}. Although it is already
shown that the Green-Kubo type formula can be used reasonably as
a good approximation
\cite{Uneyama-Masubuchi-Horio-Matsumiya-Watanabe-Pathak-Roland-2009},
here we calculate the rheo-dielectric response function exactly to
investigate the model properties precisely.
Recently, Baiesi, Maes and Wynants
\cite{Baiesi-Maes-Wynants-2009,Baiesi-Maes-Wynants-2009a} derived a
linear response formula in nonequilibrium states. Their formula does not
involve any unclear approximations, and it can be
applied to various nonequilibrium systems. Moreover, it is expressed as
a simple form which enables us to evaluate linear response functions
easily.
We show a simple derivation of the Baiesi-Maes-Wynants formula in
Appendix
\ref{derivation_of_the_baiesi_maes_wynants_formula_by_the_path_integral_formalism}.
The Baiesi-Maes-Wynants formula gives the following form as the
rheo-dielectric response function.
\begin{equation}
 \varphi(\dot{\gamma},t - t') = \frac{4 \pi \nu_{0} \tilde{\mu}^{2}}{k_{B} T}
  \bigg[ \frac{1}{2} \frac{d}{dt'} \langle R_{y}(t) R_{y}(t')
   \rangle_{\text{ss}}
   + \frac{1}{2} \bigg\langle R_{y}(t) \frac{1}{\zeta_{0}}
   \frac{\partial \mathcal{F}(\bm{R}(t'))}{\partial R_{y}(t')}
   \bigg\rangle_{\text{ss}}
   \bigg]
\end{equation}
After straightforward calculations, finally the rheo-dielectric response
function becomes a single Debye type decay function as follows.
\begin{equation}
 \label{rheo_dielectric_response_function}
 \varphi(\dot{\gamma},t) = \Delta\varepsilon_{0}
   \frac{1}{\tau_{0}} e^{- t / \tau_{0}}
\end{equation}
Now we find that the rheo-dielectric response function
\eqref{rheo_dielectric_response_function} is independent of shear
rate $\dot{\gamma}$, and thus it coincides with the equilibrium dielectric
response function, eq \eqref{dielectric_response_function_equilibrium}. Then the rheo-dielectric intensity
$\Delta\varepsilon(\dot{\gamma})$ and the rheo-dielectric relaxation
time $\tau_{\varepsilon}(\dot{\gamma})$
simply become
\begin{align}
 & \label{rheo_dielectric_intensity}
 \Delta\varepsilon(\dot{\gamma}) = \Delta\varepsilon_{0} = \frac{4 \pi \nu_{0} \tilde{\mu}^{2}
  \bar{R}^{2}}{3 k_{B} T} \\
 & \tau_{\varepsilon}(\dot{\gamma}) = \tau_{0}
\end{align}
Both are equivalent to equilibrium forms. This result is consistent with
experimental data \cite{Watanabe-Ishida-Matsumiya-2002,Watanabe-Matsumiya-Inoue-2003,Watanabe-Matsumiya-Inoue-2005}.
Strictly speaking, the term ``relaxation'' should be used to describe
``an approach'' from a nonequilibrium state to an equilibrium state. But in this work, for
convenience, we also call ``an approach'' to a perturbed state to a
nonequilibrium steady state as ``relaxation''.

Experimentally, it is convenient to use in the frequency domain
expression (Fourier transform) of the
rheo-dielectric expression,
rather than the time domain expression
\eqref{rheo_dielectric_response_function} \cite{Kubo-Toada-Hashitsume-book}. The real and imaginary parts
of Fourier transform of \eqref{rheo_dielectric_response_function} become
as $\varepsilon'(\dot{\gamma},\omega) - \varepsilon_{\infty} =
 \Delta\varepsilon_{0} / [1 + (\omega \tau_{0})^{2}]$ and 
$\varepsilon''(\dot{\gamma},\omega) = \Delta\varepsilon_{0} \omega
\tau_{0} / [1 + (\omega \tau_{0})^{2}]$.

These results are rather trivial, because the $y$-component of the
Langevin equation, eq \eqref{langevin_equation_under_shear_y}, is closed
and contains only $R_{y}(t)$ (no $R_{x}$ and $R_{z}$ contamination) and thus it is
independent of the shear rate. If dynamics of $x$- and $y$-components
are coupled, the rheo-dielectric function can be affected by the shear
rate (as shown in Appendix
\ref{weak_kinetic_coupling_between_different_directions}).
But we should be careful that even if there are no difference between
the equilibrium dielectric response and the rheo-dielectric
response under shear, the system is subjected to shear flow and thus not
in equilibrium. Actually, as we showed in the previous subsection, the
shear viscosity drastically decreased even when the rheo-dielectric
response is not affected.

\subsection{Parallel and Perpendicular Moduli}

We consider the situation where a small, time dependent
deformation is imposed to the constant-rate simple shear flow
\eqref{velocity_gradient_tensor_simple_shear}.
Experimentally, so-called
parallel or orthogonal superpositions are often utilized. In the case of
the parallel superposition, the shear rate $\dot{\gamma}$ is modulated in
time. This can be interpreted that we impose the perturbation velocity
gradient tensor which has only the $xy$-component to the system. The
excess contribution for the $xy$-component of the stress tensor is then measured.
On the other hand, in the case of the perpendicular superposition, the
$xy$-component is fixed to be constant and a small $zy$-component
velocity gradient is imposed. The $zy$-component of the stress tensor is
measured as the linear response.
In the absent of the
shear flow, these response functions coincide with the shear relaxation
modulus $G(t)$. The linear response theory gives
\begin{equation}
 \label{shear_relaxation_modulus_equilibrium_green_kubo}
 G(t - t') = \frac{1}{k_{B} T} \bigg\langle \sigma_{xy}(t)
  \left[ \frac{3 k_{B} T}{\bar{R}^{2}} R_{x}(t') R_{y}(t') \right]
  \bigg\rangle_{\text{eq}}
= \frac{9 \nu_{0} k_{B} T}{\bar{R}^{4}}
 \langle R_{x}(t) R_{y}(t)
  R_{x}(t') R_{y}(t')
 \rangle_{\text{eq}}
\end{equation}
In our model, eq \eqref{shear_relaxation_modulus_equilibrium_green_kubo}
reduces to a single Maxwell model.
\begin{equation}
 \label{shear_relaxation_modulus_equilibrium}
 G(t) = G_{0} e^{- 2 t / \tau_{0}}
\end{equation}
where we defined the characteristic modulus $G_{0}$ as $G_{0} \equiv \nu_{0} k_{B} T$.

Now we calculate the linear response of the stress tensor to the
perturbation velocity gradient tensor, under steady shear.
Unfortunately, the Baiesi-Maes-Wynants formula is limited for
perturbations which can be expressed as perturbation potentials, and
cannot be used in this case.
Here we utilize the formula given by Seifert and Speck
\cite{Seifert-Speck-2010} instead. The Seifert-Speck formula can be
applied for non-potential type perturbations.
We show a brief derivation of the formula in Appendix
\ref{derivation_of_the_baiesi_maes_wynants_formula_by_the_path_integral_formalism}.
The Seifert-Speck formula gives the following expressions for the
parallel and perpendicular moduli.
\begin{align}
 & \label{parallel_modulus_speck_seifert}
 G_{\parallel}(\dot{\gamma},t - t')
 = \frac{1}{k_{B} T} \bigg\langle \sigma_{xy}(\bm{R}(t))
 \left[R_{y}(t') \frac{1}{\Lambda_{xx}(\dot{\gamma})}
 \left[ \frac{dR_{x}(t')}{dt'} + \Lambda_{xx}(\dot{\gamma})
 \frac{\partial \mathcal{F}(\bm{R}(t'))}{\partial R_{x}(t')}
 - \dot{\gamma} R_{y}(t') \right] \right]
 \bigg\rangle_{\text{ss}} \\
 & \label{perpendicular_modulus_speck_seifert}
 G_{\perp}(\dot{\gamma},t - t')
 = \frac{1}{k_{B} T} \bigg\langle \sigma_{zy}(\bm{R}(t))
 \left[R_{y}(t') \frac{1}{\Lambda_{zz}(\dot{\gamma})}
 \left[ \frac{dR_{z}(t')}{dt'} + \Lambda_{zz}(\dot{\gamma})
 \frac{\partial \mathcal{F}(\bm{R}(t'))}{\partial R_{z}(t')}
\right] \right]
 \bigg\rangle_{\text{ss}}
\end{align}
After straightforward calculations, we find eqs \eqref{parallel_modulus_speck_seifert} and
\eqref{perpendicular_modulus_speck_seifert} reduce to simple Maxwellian
forms as follows
\begin{align}
 & \label{parallel_modulus}
 G_{\parallel}(\dot{\gamma},t)
  = G_{0} e^{- (1 + \tilde{\lambda}) t / \tau_{0}} \\
 & \label{perpendicular_modulus}
 G_{\perp}(\dot{\gamma},t)
  = G_{0} e^{- 2 t / \tau_{0}}
\end{align}
We find that the parallel modulus \eqref{parallel_modulus} differs from
the equilibrium shear relaxation modulus $G(t)$ while the perpendicular modulus \eqref{perpendicular_modulus} is
identical to $G(t)$.
Both $G_{\parallel}(\dot{\gamma},t)$ and $G_{\perp}(\dot{\gamma},t)$ approach $G_{0}$ at the short time
limit.
The parallel and perpendicular relaxation times under shear are given by
\begin{align}
 & \tau_{\parallel}(\dot{\gamma})
  = \frac{\tau_{0}}{1 + \tilde{\lambda}} \\
 & \tau_{\perp}(\dot{\gamma})
 = \frac{\tau_{0}}{2}
\end{align}
The parallel relaxation time depends on the shear rate in the same way as
the shear viscosity.
We find that, in our model, the effect of the shear
rate to the parallel relaxation modulus is observed only as the
acceleration of the relaxation time. As we increase the shear rate, the
parallel relaxation
time decreases. This is in contrast to the rheo-dielectric function
where we have no effect of the shear rate.
Theoretically, the parallel modulus depends on the shear
rate because it involves the correlation function of the $x$-component.

\begin{figure}[h!]
\includegraphics{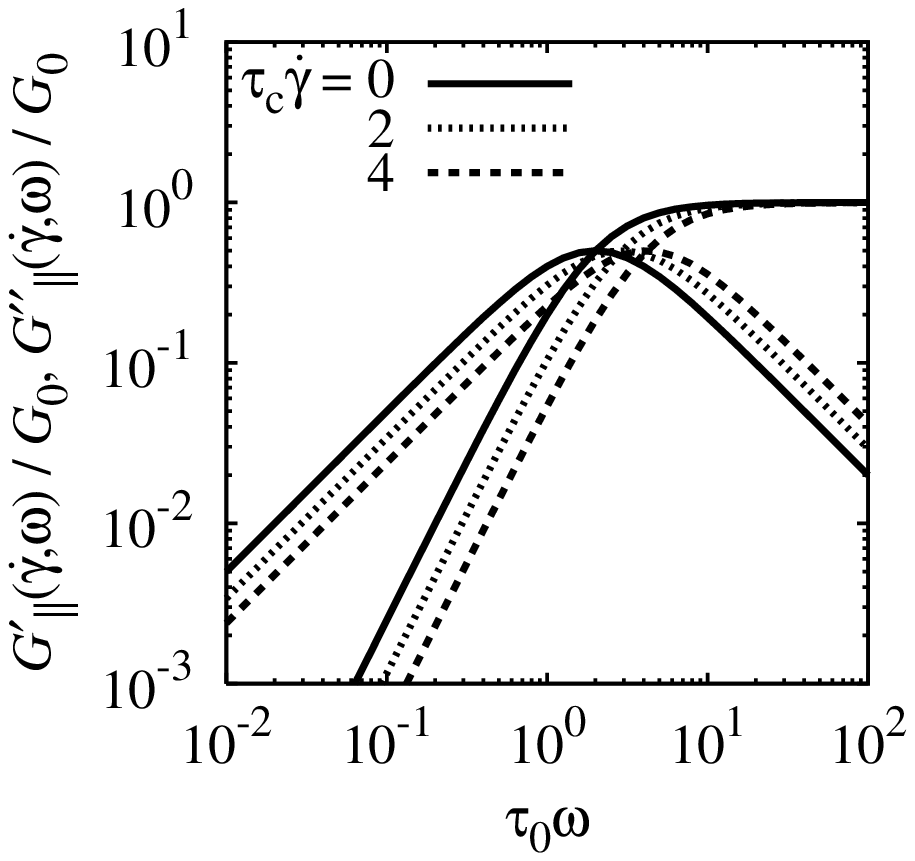}
\caption{\label{parallel_moduli_power_law}}
\end{figure}

As in the case of the rheo-dielectric response, it is convenient to use the
Fourier transformed response functions in the frequency domain \cite{Doi-Edwards-book}.
From eqs \eqref{parallel_modulus} and \eqref{perpendicular_modulus},
the real and imaginary parts (storage and loss) moduli become
$ G_{\parallel}'(\dot{\gamma},\omega) = G_{0} (\omega
\tau_{\parallel}(\dot{\gamma}))^{2} / [1 + (\omega
\tau_{\parallel}(\dot{\gamma}))^{2}] $,
$G_{\parallel}''(\dot{\gamma},\omega)
  = G_{0} \omega \tau_{\parallel}(\dot{\gamma}) / 
[1 + (\omega \tau_{\parallel}(\dot{\gamma}))^{2}]$,
$ G_{\perp}'(\dot{\gamma},\omega)
  = G_{0} (\omega \tau_{0} / 2)^{2} / [1 + (\omega \tau_{0} / 2)^{2}]$, and
$G_{\perp}''(\dot{\gamma},\omega)
  = G_{0} (\omega \tau_{0} / 2) / [1 + (\omega \tau_{0} / 2)^{2}]$.
We show parallel storage and loss moduli for several values of
$\tau_{c}\dot{\gamma}$ with the power law model
(eq \eqref{dimensionless_mobility_factor_power_law}) in FIG. \ref{parallel_moduli_power_law}.

Experimentally, both parallel and perpendicular relaxation times,
$\tau_{\parallel}$ and $\tau_{\perp}$,
decrease as the shear rate increases, and the decrease is more
systematic for $\tau_{\parallel}$ \cite{Vermant-Walker-Moldenaers-Mewis-1998}. Our model predicts the decrease of the parallel relaxation time,
which is consistent with the experimental data. On the other hand, the
perpendicular relaxation time in our model is independent of the shear
rate. This is because the dynamics of $y$- and $z$-components are not
coupled to one of $x$-component.
Thus we conclude that our model can
reproduce the parallel modulus qualitatively but cannot reproduce the
perpendicular modulus. (This is due to the oversimplification.)
Although the parallel or perpendicular moduli have been analyzed or explained
mainly on the basis of constitutive equation models \cite{Macdonald-Bird-1966,Tanner-1968,Yamamoto-1971,Vermant-Walker-Moldenaers-Mewis-1998}, as far as we know, there
is no analysis/explanation on the basis of the linear response theory.

\section{Discussions}

\subsection{Anisotropic Mobility Tensor}

Although our model is too simple to apply practical analyses, it still
specifies some characteristic features of polymer dynamics under shear.
We may comment that the anisotropic mobility tensor model proposed in
this work can be understood as a simplified model studied in a
previous theoretical work on the rheo-dielectric response
\cite{Uneyama-Masubuchi-Horio-Matsumiya-Watanabe-Pathak-Roland-2009}. In
the previous work, we have
studied the rheo-dielectric response function of a linearized Langevin
equation for an entangled polymer.
The anisotropic mobility tensor model
is linear and has the same properties as the linearized Langevin
equation model. (The expression of the rheo-dielectric response
function is not affected significantly by the shear rate.)
Our anisotropic mobility model reinforces the validity of the linearized
Langevin equation model. Further, we expect that the linearized Langevin
model will reproduce similar properties as the anisotropic mobility
model (such as the acceleration of the parallel relaxation time).

To model the dynamics of entangled polymers under fast shear, currently the convective
constraint release (CCR) model \cite{Marrucci-1996} is widely employed. The CCR model claims
that the effective relaxation time is accelerated under shear
flow, due to the enhancement of the constraint release. Theories or simulations which take account the CCR effect achieved
success to explain or reproduce rheological behaviour of entangled
polymers under shear \cite{Marrucci-1996,Ianniruberto-Marrucci-1996,Mead-Larson-Doi-1998,Ianniruberto-Marrucci-2001,Masubuchi-Takimoto-Koyama-Ianniruberto-Greco-Marrucci-2001,Graham-Likhtman-McLeish-Milner-2003}.
While the CCR model originally gives the expression for the
relaxation time, here we interpret the modification of the relaxation
time as the modification of the mobility (or the friction coefficient).
Then the CCR model is interpreted as the shear rate dependent
mobility model.
The expression of the CCR mobility becomes as follows.
\begin{equation}
 \label{mobility_tensor_ccr}
 \bm{\Lambda}_{\text{CCR}}(\dot{\gamma})
  = \left[ \frac{1}{\zeta_{0}} + \frac{\beta_{\text{CCR}}}{k_{B} T} \bm{\kappa} : \langle \bm{R} \bm{R}
   \rangle \right] \bm{1}
  = \left[ \frac{1}{\zeta_{0}} + \frac{\beta_{\text{CCR}}}{k_{B} T} \dot{\gamma} \langle R_{x} R_{y}
   \rangle \right] \bm{1}
\end{equation}
where $\beta_{\text{CCR}}$ is a positive constant of the order of
unity and $\langle \dots \rangle$ represents the statistical (ensemble) average. In the steady state, $\langle \bm{R} \bm{R} \rangle$ can be
replaced by the steady state average $\langle \bm{R} \bm{R}
\rangle_{\text{ss}}$, which is determined self-consistently.

It is obvious that
the CCR mobility \eqref{mobility_tensor_ccr} is isotropic, and thus accelerates chain motion in all
directions. As a
result, linear response functions such as rheo-dielectric response
function or the parallel modulus are affected by the shear flow.
Although experimental data of parallel moduli can be
reproduced well by the CCR model
\cite{Somma-Valentino-Titomanlio-Ianniruberto-2007,Boukany-Wang-2009,Li-Wang-2010},
the rheo-dielectric response data
\cite{Matsumiya-Watanabe-Inoue-Osaki-Yao-1998,Watanabe-Ishida-Matsumiya-2002,Watanabe-Matsumiya-Inoue-2003,Watanabe-Matsumiya-Inoue-2005}
cannot be reproduced.
In other words, the isotropic acceleration by the CCR is not consistent with the
experimental data for the rheo-dielectric responses.
As we have shown in the previous section, the anisotropic model can naturally overcome
this difficulty.
Therefore we consider that the CCR model needs to be modified to
reproduce the anisotropic chain motion (or the anisotropic acceleration).
For example, we may introduce a phenomenological
anisotropic relaxation time tensor instead of
a scalar relaxation time. Then the resulting model will reproduce
the rheo-dielectric function which is insensitive to the shear rate, as well as the shear thinning.
One simple possible modification is shown in Appendix
\ref{anisotropic_version_of_the_convective_constraint_release_model}.

To compare the CCR model with our anisotropic mobility model from a
different aspect, here we consider a general form for the mobility tensor.
We consider the mobility tensor under shear flow in general flow and
gradient directions.
From the symmetry of the Langevin equation under rotational
transform, the mobility tensor $\bm{\Lambda}$ should be invariant for
rotational transform.
That means, we can expand $\bm{\Lambda}$ into a power series of scalar
and tensor invariants if the shear rate is not high.
Then, up to the second order in $\dot{\gamma}$, the expansion form of
$\bm{\Lambda}$ is given as
\begin{equation}
 \label{mobility_tensor_general_expansion}
 \bm{\Lambda}(\bm{\kappa}) = \frac{1}{\zeta_{0}}
  \left[ \bm{1}
   + \tilde{L}_{1} (\bm{\kappa} + \bm{\kappa}^{t})
   + \tilde{L}_{2} \operatorname{tr} (\bm{\kappa} \cdot \bm{\kappa}^{t})
   \bm{1}
   + \tilde{L}_{3} \bm{\kappa} \cdot \bm{\kappa}^{t}
   + \tilde{L}_{4} \bm{\kappa}^{t} \cdot \bm{\kappa}
   + O(\dot{\gamma}^{3}) \right]
\end{equation}
Here $\lbrace \tilde{L}_{i} \rbrace$ is a set of expansion
coefficients.
We find that the CCR mobility model \eqref{mobility_tensor_ccr} can be
reproduced by setting $\tilde{L}_{1} = \tilde{L}_{3} = \tilde{L}_{4} = 0$, while our anisotropic mobility
model can be reproduced by setting $\tilde{L}_{1} = \tilde{L}_{2} = \tilde{L}_{4} = 0$. (If $\dot{\gamma}$
is sufficiently small, we can set $\langle R_{x} R_{y} \rangle = \dot{\gamma}
\tau_{0} \bar{R}^{2} / 6 + O(\dot{\gamma}^{2})$
in eq
\eqref{mobility_tensor_ccr}.
If $\bm{\kappa}$ is given by eq
\eqref{velocity_gradient_tensor_simple_shear}, only the $xx$-component of $\bm{\kappa} \cdot
\bm{\kappa}^{t}$ is nonzero.) Therefore both the CCR model and our model
are allowed from the symmetry argument. The mobility tensor should be
modelled so that the resulting dynamics reproduces required properties
(such as the insensitivity of the rheo-dielectric response function to
the shear rate). We may employ a different mobility tensor model which
is reduce to eq \eqref{mobility_tensor_general_expansion}. For
example, by setting $\tilde{L}_{2} = \tilde{L}_{3} = \tilde{L}_{4} = 0$,
we have a nondiagonal mobility tensor model which takes account of
the kinetic coupling effect.
(See Appendix \ref{weak_kinetic_coupling_between_different_directions}.)
Here we should note that in the conventional approach
\cite{Beris-Edwards-book}, the mobility tensor is expressed as a
function of the average conformation tensor (not as a function of the
velocity gradient tensor), and thus it is expanded into
a power series of the conformation tensor. However, from the view point
of the nonequilibrium statistical physics
\cite{McPhie-Daivis-Snook-Ennis-Evans-2001}, in principle,
the mobility tensor can depend on the velocity gradient tensor and can be
modelled as eq \eqref{mobility_tensor_model} or eq
\eqref{mobility_tensor_general_expansion}.
(We should also note that the current approach is limited for simple
shear flows, and for other flows, such as elongational flows, we need to
construct the mobility tensor model.)

One may argue that if an isotropic mobility (which can be utilized
near equilibrium) should be replaced by an anisotropic mobility for
nonequilibrium systems, then
an equilibrium free energy \eqref{free_energy_linear_elasticity} should
also be replaced by a nonequilibrium effective free energy. 
This is indeed correct in general. Nevertheless, for polymeric systems we can use the
equilibrium free energy \eqref{free_energy_linear_elasticity} safely
even under fast shear. This is because in polymeric systems, the
stress-optical rule \cite{Doi-Edwards-book,Inoue-Osaki-1996} is known to be hold even
under fast shear.
(Although it is known that the stress-optical rule fails for some cases,
such as systems under fast extensional
flow \cite{Sridhar-Nguyen-Fuller-2000,Luap-Muller-Schweizer-Venerus-2005}, it is valid under the current situation.)
The stress-optical rule relates the chain conformation and the force
exerted by the chain, and holds only when the force is proportional to
the bond vector.
This gives the linear entropic elasticity model for the free
energy, and as long as the stress-optical rule holds we can
justify the use of the equilibrium free energy
\eqref{free_energy_linear_elasticity} even under fast shear.

At the end of this subsection, we may comment on one assumption used
in our model.
We have assumed the fluctuation-dissipation like relation
for the thermal noise (eq
\eqref{fluctuation_dissipation_relation_2nd_moment}). Such a relation does
not necessarily hold in the nonequilibrium states, and thus we can
employ nondiagonal mobility tensor model. From
the view point of the linear response theory, whether the mobility
tensor is diagonal or not is not so essential (see Appendix
\ref{derivation_of_the_baiesi_maes_wynants_formula_by_the_path_integral_formalism}).
Although we do
not discuss further in detail about the nondiagonal mobility tensor
because it is beyond the scope of this work, we expect it may be required
to describe polymer dynamics precisely under general flow conditions.

\subsection{Entropy Production and Steady State Probability Current}

We have derived expressions for several linear response functions.
In several pieces of
theoretical work, the violation of the fluctuation-dissipation relation is
interpreted as the entropy production rate \cite{Harada-Sasa-2005,Speck-Seifert-2006}, the steady state probability
current \cite{Chetrite-Falkovich-Gawedzki-2008,Chetrite-Gawedzki-2009,Ohta-Ohkuma-2008}, or other related physical quantities.
Here we calculate the entropy production rate for our model
and investigate how it is related to the response functions.

\begin{figure}[h!]
\includegraphics{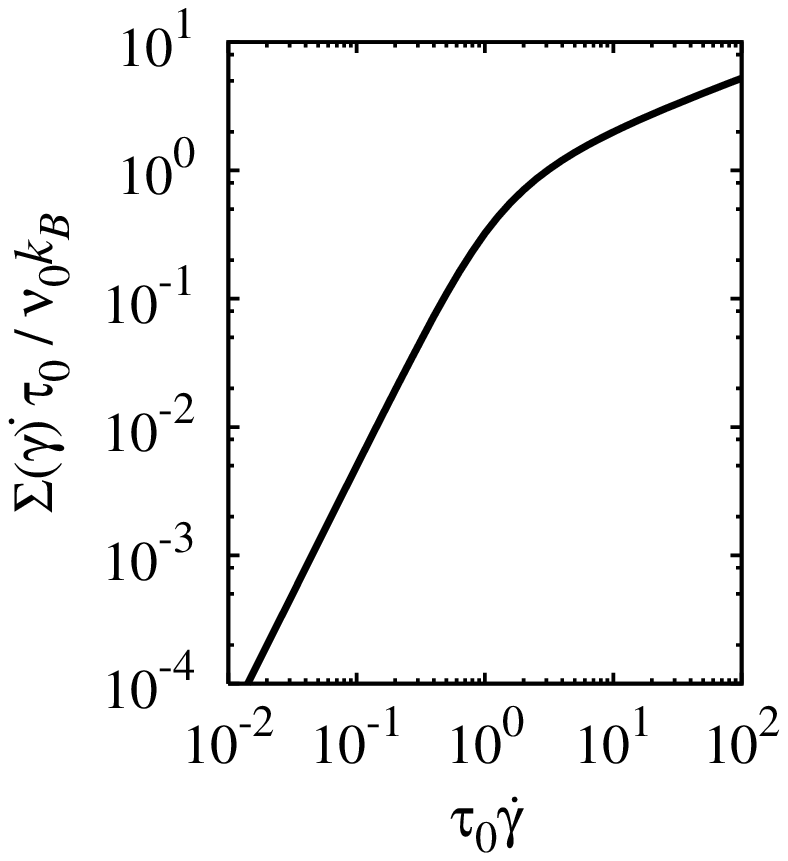}
\caption{\label{entropy_production_rate_power_law}}
\end{figure}

Following the standard definition
\cite{Lebowitz-Spohn-1999,Qian-2001,Maes-Netocny-Wynants-2008} we define the steady
state entropy production rate per unit volume as follows.
\begin{equation}
 \label{entropy_production_rate_definition}
  \Sigma(\dot{\gamma})
  \equiv \frac{\nu_{0}}{T} \int d\bm{r} \,
  \frac{\bm{J}_{\text{ss}}(\bm{r}) \cdot \bm{\Lambda}^{-1}(\dot{\gamma})
  \cdot \bm{J}_{\text{ss}}(\bm{r})}{P_{\text{ss}}(\bm{r})} 
\end{equation}
where $\bm{J}_{\text{ss}}(\bm{r})$ is the steady state probability
current defined as
\begin{equation}
 \label{steady_state_probability_current_definition}
  \bm{J}_{\text{ss}}(\bm{r})
   \equiv
  - \bm{\Lambda}(\dot{\gamma}) \cdot
  \left[ \frac{\partial \mathcal{F}(\bm{r})}{\partial \bm{r}}
   P_{\text{ss}}(\bm{r}) + k_{B} T \frac{\partial P_{\text{ss}}(\bm{r})}{\partial \bm{r}} \right]
  + \bm{\kappa} \cdot \bm{r} P_{\text{ss}}(\bm{r})
\end{equation}
The steady state probability current
\eqref{steady_state_probability_current_definition} satisfies the steady
state condition, $(\partial / \partial \bm{r}) \cdot \bm{J}_{\text{ss}}(\bm{r})
= 0$.
After the straightforward calculation, we have the explicit expression
for the steady state entropy production rate in our model.
\begin{equation}
 \label{entropy_production_rate}
 \begin{split}
  \Sigma(\dot{\gamma})
  & = \frac{\nu_{0}}{T} \bm{\Lambda}(\dot{\gamma}) : \int d\bm{r} \,    P_{\text{ss}}(\bm{r})
\bigg[ \frac{3 k_{B} T}{\bar{R}^{2}} \bm{r}
  - \bm{\Lambda}^{-1}(\dot{\gamma}) \cdot \bm{\kappa} \cdot \bm{r} \bigg]
  \bigg[
  \frac{3 k_{B} T}{\bar{R}^{2}}  \bm{r}
  -  \bm{\Lambda}^{-1}(\dot{\gamma}) \cdot 
  \bm{\kappa} \cdot \bm{r}
  \bigg]  - \frac{3 \nu_{0} k_{B}^{2} T}{\bar{R}^{2}}
  \bm{\Lambda}(\dot{\gamma}) : \bm{1} \\
  & = \frac{\nu_{0} \zeta_{0}^{2} \tilde{\lambda}}{T}
  \bigg( \frac{1}{\tau_{0}^{2}} C_{xx}
  - \frac{ 2 \dot{\gamma}}{\tau_{0} \tilde{\lambda}}
  C_{xy}
  + \frac{ \dot{\gamma}^{2}}{\tilde{\lambda}^{2}} C_{yy} \bigg)
 + \frac{\nu_{0} \zeta_{0}^{2}}{T \tau_{0}^{2}}
  ( C_{yy} + C_{zz} ) - \frac{\nu_{0} k_{B}}{\tau_{0}}
  ( \tilde{\lambda} + 2 ) \\
  & = \frac{\nu_{0} k_{B}}{\tau_{0}} \frac{(\tau_{0} \dot{\gamma})^{2}}{\tilde{\lambda}(1 + \tilde{\lambda})}
 \end{split}
\end{equation}
As expected, the entropy production rate
\eqref{entropy_production_rate} is the even function of
$\dot{\gamma}$, and it is nonzero unless $\dot{\gamma} = 0$.
We show the entropy production rate for the power law type model
($\alpha = 9 / 11$ and $\tau_{c} = \tau_{0}$) in
FIG. \ref{entropy_production_rate_power_law}. We can employ another
definition for the entropy production rate, $\Sigma \equiv \dot{\gamma} \langle
\sigma_{xy} \rangle_{\text{ss}} / T$
\cite{Evans-Searles-2002}. This gives a slightly different form from eq
\eqref{entropy_production_rate}, but the result is qualitatively the same.

Some nonequilibrium linear response theories state that, the entropy
production rate (which may be interpreted as the distance from
equilibrium) is related to the violation of the Green-Kubo type response
formulae. We call such a picture as the entropy production picture.
In the entropy production picture, one expects that if the system is not in
equilibrium, there should be the correction terms in the linear response
formulae which is directly related to the entropy production rate.
However, as we have already shown, the rheo-dielectric response function
is unchanged even under shear in our model.
Besides, in the previous work \cite{Uneyama-Masubuchi-Horio-Matsumiya-Watanabe-Pathak-Roland-2009}, the Green-Kubo type
formula was shown to be approximately valid for the rheo-dielectric response.
These results mean that, some linear response functions do not change
their forms even in the nonequilibrium states. This may sound inconsistent with the entropy
production picture.
This is because what appears in linear response formulae is not the
entropy production rate itself but the derivative of the entropy
production rate with respect to an external perturbation field.

The situation may be clearer if we employ another picture, which focus on
the steady state probability current. A linear response function in
nonequilibrium steady state can be expressed as the sum of the
Green-Kubo type equilibrium form and the correction term, which involves
the probability current. We call this picture as the Lagrangian
moving frame picture \cite{Chetrite-Falkovich-Gawedzki-2008,Chetrite-Gawedzki-2009}. The Lagrangian moving frame picture
gives the following response formula.
\begin{equation}
 \label{linear_response_formula_lagrangian_moving_frame}
 \mathcal{R}_{AB}(t - t')
 = \frac{1}{k_{B} T} \frac{d}{dt'} \langle A(t) B(t') \rangle_{\text{ss}}
 - \frac{1}{k_{B} T} \bigg \langle A(t) \bm{v}_{\text{ss}}(\bm{R}(t')) \cdot
 \frac{\partial B(t')}{\partial \bm{R}(t')} \bigg\rangle_{\text{ss}}
\end{equation}
where $\bm{v}_{\text{ss}}(\bm{r})$ is the mean steady state streaming
velocity defined as $\bm{v}_{\text{ss}}(\bm{r}) \equiv \bm{J}_{\text{ss}}(\bm{r}) / P_{\text{ss}}(\bm{r})$.
The second term in the right hand side of eq
\eqref{linear_response_formula_lagrangian_moving_frame} is the
correction term due to the nonzero steady state probability current.
The correction term can be zero even if $\bm{J}_{\text{ss}}(\bm{r})
\neq 0$. In the case of the rheo-dielectric response in our
model, the correction term is exactly equal to zero. This can be shown
straightforwardly.
\begin{equation}
 \begin{split}
  \bigg\langle [4 \pi \nu_{0} \tilde{\mu} R_{y}(t)]
  \bm{v}_{\text{ss}}(\bm{R}(t')) \cdot \frac{\partial [\tilde{\mu} R_{y}(t')]}{\partial
  \bm{R}(t')} \bigg\rangle_{\text{ss}}
  & \propto \bigg\langle R_{y}(t)
  \left[ R_{x}(t') - \frac{\tau_{0} \dot{\gamma}}{1 + \tilde{\lambda}}
   R_{y}(t')
  \right] \bigg\rangle_{\text{ss}} \\
  & = e^{- (t - t') / \tau_{0}}
  \left[  \langle R_{y} R_{x} \rangle_{\text{ss}}
  - \frac{\tau_{0} \dot{\gamma}}{1 + \tilde{\lambda}}
  \langle R_{y}^{2} \rangle_{\text{ss}}
  \right] \\
  & = 0
 \end{split}
\end{equation}
Moreover, it is
quite difficult to separate an experimentally measured response function
into two terms as eq
\eqref{linear_response_formula_lagrangian_moving_frame}. This is
because the first term in the right hand side of eq
\eqref{linear_response_formula_lagrangian_moving_frame} (the
Green-Kubo type term) can also depend on the shear rate.

From the results and discussions above, we consider that {\em even if} we measure 
the rheo-dielectric function and the entropy production rate of the same system
under shear simultaneously, we will not be able to verify
the violation of the Green-Kubo type formula precisely.
Thus we consider that the entropy production or the Lagrangian moving
frame pictures are not so useful to analyze rheo-dielectric responses or
other linear responses under shear.
To investigate dynamics of
polymer chains under shear in detail, we consider it is better to
measure other linear
response functions, such as the parallel modulus, instead of the entropy
production rate (or the corresponding heat flow). Combination of several
linear response functions will provide us detail information about the
dynamics of polymer chain \cite{Watanabe-1999,Matsumiya-Watanabe-Osaki-2000,Watanabe-Matsumiya-Osaki-2000,Matsumiya-Watanabe-2001,Watanabe-Matsumiya-Inoue-2002}.

To be fair, we should
mention that for microscopic systems (such as a colloid particle driven by
an optical trap \cite{Blickle-Speck-Lutz-Seifert-Bechinger-2007,GomezSolano-Petrosyan-Ciliberto-Chetrite-Gawedzki-2009}) the
entropy production rate or related quantities can be utilized
successfully to characterize the nonequilibrium features.
For carefully designed microscopic systems we can measure correlation functions or the
steady state probability current directly by microscopes. These physical
quantities are essential in the entropy production or Lagrangian moving
frame pictures.
But in macroscopic systems, it is quite difficult or practically impossible to
measure several physical quantities.
Thus a different approach for macroscopic systems is naturally required.

\subsection{Dependence on Architecture of Polymers}

We have shown that our anisotropic mobility tensor model can reproduce
rheo-dielectric response behavior qualitatively. That is, the
rheo-dielectric response functions of linear polymers are insensitive to
the shear rate. However, rheo-dielectric response functions of star
polymers are reported to slightly depend on the shear rate \cite{Watanabe-Ishida-Matsumiya-2002}.
This cannot be explained by our model. In this subsection, we consider
why our model fails to describe star polymers and possible ways to
improve the model.

Arsac et al \cite{Arsac-Carrot-Guillet-Revenu-1994} fit experimental
rheology data to the JS model and determined the slip factors (the
fitting parameters in the JS model).
They found that the slip factors for linear polymers are nearly independent of the flow regime
(transient or steady state) or the molecular weight
distribution.
This means that the JS model can reproduce dynamics of
linear polymers in spite of its very simple form.
However, for branched polymers the slip parameters depend on various
factors.
We may say that the dynamics of entangled linear
polymers is rather simple, in a sense.
Thus we consider that the
dynamics of star polymers cannot be described well by a simple model
like the JS model.

Matsumiya, Watanabe and coworkers
\cite{Matsumiya-Watanabe-Osaki-2000,Watanabe-Matsumiya-Osaki-2000,Matsumiya-Watanabe-2001,Watanabe-Matsumiya-Inoue-2002}
measured and analyzed
the linear viscoelasticities and dielectric responses of entangled
linear and star polymers. They quantitatively tested the dynamic tube
dilation (DTD) model \cite{Marrucci-1985} for linear and star polymers.
They reported that the simple DTD model explains
the experimental data for linear polymers well
\cite{Matsumiya-Watanabe-Osaki-2000}, while it fails
for star polymers
\cite{Watanabe-Matsumiya-Osaki-2000,Matsumiya-Watanabe-2001,Watanabe-Matsumiya-Inoue-2002}.
This failure is attributed to
the overestimate of the equilibration by
the constraint release at the long time region
\cite{Watanabe-Ishida-Matsumiya-Inoue-2004,Watanabe-Sawada-Matsumiya-2006,Watanabe-2008},
or the events that newly created entanglements push out the old
entanglements toward chain ends \cite{Shanbhag-Larson-Takimoto-Doi-2001}.
These experimental results indicate that 
a rather simple model can describe the dynamic behavior of entangled linear polymers
but not of star polymers.

Thus we expect linear polymers have rather simple dynamic properties
even under fast flow while star polymers do not.
This implies that our simple anisotropic mobility model will be able to
describe dynamics of linear polymers qualitatively well.
On the other hand, the dynamics of branched polymers (including star
polymers) is much more complicated compared with linear polymers and too
simplified models (like ours) cannot explain dynamics of branched polymers.
Then we can conclude that our model should not be applied for star
polymers directly, and some modifications are required.

There are several possible ways to improve our model. For example, we
can employ the nondiagonal mobility tensor model, which represents the
kinetic coupling between dynamics of different directions (the kinetic
coupling model). Such a model
can reproduce the dependence of the rheo-dielectric response to the
shear rate to some extent.
We show the rheo-dielectric response function for a simple and weak kinetic
coupling model in Appendix
\ref{weak_kinetic_coupling_between_different_directions}.
Similarly, we can employ the conformation dependent mobility tensor model
\cite{Curtiss-Bird-1980,Curtiss-Bird-1980a,Giesekus-1982,PhanThien-Manero-Leal-1984,Baxandall-1987,Biller-Petruccione-1988}.
The conformation dependent mobility kinetically couples the dynamics for different
directions, and will give similar results as simple kinetic
coupling models.

Another possible way is to employ a fine scale
description such as the full bead-spring type model with topological
constraints \cite{Likhtman-2005}. Integrating our
anisotropic mobility model into bead-spring type models will allow us
to study the dependence of the rheo-dielectric behavior on the polymer
architecture. Anyway, the rheo-dielectric behavior and dynamics of
entangled star polymers are still not fully understood and further
theoretical developments are required.

\section{Conclusion}

In this work, we proposed the anisotropic mobility model for polymers
under shear. In the anisotropic mobility model, the mobility tensor or
the friction coefficient tensor becomes anisotropic and dependent on the
shear rate. The anisotropic mobility tensor model is consistent with
NEMD simulation results and thus we expect that our model captures
qualitative and essential nature of polymer dynamics under shear.

We calculated the shear viscosity, the rheo-dielectric
response function, or the parallel and perpendicular moduli. Our model
gives the rheo-dielectric function which is independent of the shear
rate, even when the shear rate is sufficiently high and shear thinning
is exhibited. This is qualitatively consistent with the experimental results.
Our model gives the parallel relaxation time which decreases with increasing the shear
rate. This is also qualitatively consistent with the experimental results.
Of course, the shear-rate insensitive rheo-dielectric relaxation and
acceleration of the parallel relaxation observed in experiments may
result from not only the anisotropic mobility but also from other
factors (such as full DTD in the linear response regime). However, the
current study demonstrates that the anisotropic mobility could play an
important role in the relaxation processes.

To examine the properties of our model in detail,
we compared our model with other models or theories. We compare our
model with the CCR model.
Both our model and the CCR model accelerate the dynamics of polymers
under shear. Our model accelerate the dynamics anisotropically while the
CCR model accelerate the dynamics isotropically. Judging from the
experimental results, we consider our model is more suitable to describe
the dynamics of polymers under shear.

We also compared our result with the recent linear response theories for
nonequilibrium systems. Although the entropy production rate or the
steady state probability current are widely utilized in recent models, we
showed that they are not so useful to analyze or understand the rheo-dielectric
response function. 
We consider that the combination of several
different linear response functions will be reasonable to
investigate polymer dynamics under shear.

Although our model can explain the essential feature of linear polymers under
shear, it should be improved or modified further.
For example, our model can not explain the experimental results for star
polymers under shear. We did not explicitly consider the effect of
entanglements, which will be important for star polymers.
The integration of our anisotropic mobility model into fine scale models
or the generalization of our model to general flow conditions is
considered to be an interesting subject of future work.

\section*{Acknowledgment}

This work was supported by Grant-in-Aid (KAKENHI) for Young
Scientists B 22740273, and by the Research Fellowships of the
Japan Society for the Promotion of Science for Young Scientists.

\appendix

\section{Constitutive Equation for Anisotropic Mobility Model}
\label{constitutive_equation_for_anisotropic_mobility_model}
In this appendix, we derive the constitutive equation from the
anisotropic mobility model and compare it with some conventional models.
For simplicity, we assume that the system is homogeneous in this
appendix. (The generalization for inhomogeneous systems is straightforward.)
We consider to express the constitutive equation as the dynamic equation
for the following time-dependent conformation tensor.
\begin{equation}
 \bm{C}(t) \equiv \langle \bm{R}(t) \bm{R}(t) \rangle
  = \int d\bm{r} \, \bm{r} \bm{r} P(\bm{r},t)
\end{equation}
The time evolution equation for the conformation tensor can be easily calculated
from the Fokker-Planck equation \eqref{fokker_planck_equation_under_shear}.
After a straightforward calculation, we have the following constitutive
equation for the anisotropic mobility model (eq
\eqref{langevin_equation_under_shear} together with eq
\eqref{mobility_tensor_model} or eq \eqref{mobility_tensor_general_expansion}).
\begin{equation}
 \label{constitutive_equation_anisotropic_mobility}
 \overset{\nabla}{\bm{C}}(t) =
  - \frac{3 k_{B} T}{\bar{R}^{2}}
  [\bm{C} \cdot \bm{\Lambda}(\bm{\kappa})
  + \bm{\Lambda}(\bm{\kappa}) \cdot \bm{C} ]
  + 2 k_{B} T \bm{\Lambda}(\bm{\kappa})
\end{equation}
where we defined the upper-convected derivative as $\overset{\nabla}{\bm{C}} \equiv d\bm{C} / dt
- \bm{\kappa} \cdot \bm{C} - \bm{C} \cdot \bm{\kappa}^{t}$.
It should be noticed that the anisotropic mobility model is designed
around the steady state under simple shear and thus
the corresponding constitutive equation
\eqref{constitutive_equation_anisotropic_mobility} is also applicable
around the steady state. (It is not suitable to calculate, for example,
the start-up shear flow. To study such transient phenomena, we will
need to describe the time evolution of the the mobility tensor
$\bm{\Lambda}$ explicitly.)

It is informative to compare eq
\eqref{constitutive_equation_anisotropic_mobility} with other
constitutive equation models. Although there are many constitutive
equation models for polymeric systems \cite{Beris-Edwards-book}, for the sake of simplicity, here
we limit ourselves to rather simple models. One of the simplest
constitutive equation models with anisotropic mobilities is the Giesekus
model.
The Giesekus model \cite{Giesekus-1982} employs the
conformation tensor dependent mobility, whereas the anisotropic mobility
model employs the
mobility which does not depend on the conformation tensor.
The Giesekus constitutive equation can be expressed as follows.
\begin{equation}
 \label{constitutive_equation_giesekus}
 \overset{\nabla}{\bm{C}}(t) =
  \frac{2}{\zeta_{0}}
  \left[ (1 - \alpha) \bm{1} + \frac{3 \alpha}{\bar{R}^{2}} \bm{C} \right] \cdot
  \left[ - \frac{3 k_{B} T}{\bar{R}^{2}} \bm{C}
  + k_{B} T \bm{1} \right]
\end{equation}
where $\alpha$ is a phenomenological constant ($0 \le \alpha
\le 1$).
(The Giesekus model corresponds to the pre-averaged and linearized
Curtiss-Bird model
\cite{Curtiss-Bird-1980,Curtiss-Bird-1980a,Giesekus-1982}.)
Eq \eqref{constitutive_equation_giesekus} can be obtained by replacing
$\bm{\Lambda}(\bm{\kappa})$ in eq
\eqref{constitutive_equation_anisotropic_mobility} by
$[(1 - \alpha) \bm{1} + (3 \alpha / \bar{R}^{2}) \bm{C}] / \zeta_{0}$.
Inversely we can replace $[(1 - \alpha) \bm{1} + (3 \alpha /
\bar{R}^{2}) \bm{C}] / \zeta_{0}$ by $\bm{\Lambda}(\bm{\kappa})$ to obtain
eq \eqref{constitutive_equation_anisotropic_mobility} from eq \eqref{constitutive_equation_giesekus}.
One may interpret
the anisotropic mobility model as the conventional anisotropic tensor
model with some approximations, for example, the pre-averaging around the
steady-state.
(We note that it is not simple to analyze eq
\eqref{constitutive_equation_giesekus} due to its nonlinearity, unlike
eq \eqref{constitutive_equation_anisotropic_mobility}. The conformation
tensor independent mobility tensor makes analyses in the main text simple and tractable.)

Another simple constitutive equation model is the Johnson-Segalman (JS)
model \cite{Johnson-Segalman-1977}.
The JS model employs the Gordon-Schowalter derivative
\cite{Gordon-Showalter-1972} to
produce the non-affine motion, which can be interpreted as the slippage effect.
Here we rewrite the JS model
by using the upper-convected derivative, to compare it with eq
\eqref{constitutive_equation_anisotropic_mobility}.
\begin{equation}
 \label{constitutive_equation_johnson_segalman}
 \overset{\nabla}{\bm{C}}(t) =
 \frac{2}{\zeta_{0}} \left[ - \frac{3 k_{B} T}{\bar{R}^{2}} \bm{C}
  + k_{B} T \bm{1} \right]
  + \frac{a - 1}{2}
  [ (\bm{\kappa} + \bm{\kappa}^{t}) \cdot \bm{C}
  + \bm{C} \cdot (\bm{\kappa} + \bm{\kappa}^{t}) ]
\end{equation}
Here $a$ is a phenomenological constant ($-1 \le a \le 1$), which is sometimes
called the slip parameter.
We find that the form of the JS model
\eqref{constitutive_equation_johnson_segalman} is
somehow similar to the anisotropic mobility model \eqref{constitutive_equation_anisotropic_mobility}.
The last term in the right hand side of
eq \eqref{constitutive_equation_johnson_segalman} is mathematically
similar to the first term in the right hand side of eq
\eqref{constitutive_equation_anisotropic_mobility}.
Thus we expect that the anisotropic mobility model and the JS model will show
qualitatively similar dynamical behavior in some cases. However, the
origin of that term in the JS model is the
non-affine motion (or the slippage).
Unlike the case of the Giesekus model, we cannot obtain the anisotropic
mobility model \eqref{constitutive_equation_anisotropic_mobility}
by simply replacing a part (such as the mobility tensor) in eq \eqref{constitutive_equation_johnson_segalman}.

\section{Derivation of the Baiesi-Maes-Wynants Formula by the Path Integral
Formalism}
\label{derivation_of_the_baiesi_maes_wynants_formula_by_the_path_integral_formalism}
In this appendix, we show the derivation of the Baiesi-Maes-Wynants
formula \cite{Baiesi-Maes-Wynants-2009,Baiesi-Maes-Wynants-2009a} in
steady state based on the path integral formalism.
The derivation of the linear response formula in nonequilibrium steady
state based on the path integral
formalism is first shown by Seifert and Speck \cite{Seifert-Speck-2010}.
Here we mainly follow their derivation.
It is worth noting that the path integral formalism
had already utilized by Ohta and
Ohkuma \cite{Ohta-Ohkuma-2008} to derive a similar but slightly
different formula, in prior to the Seifert-Speck theory.

The Seifert-Speck formula reduces to the Baiesi-Maes-Wynants formula if
the perturbation is expressed as an perturbation potential.
As far as we know, an explicit derivation of the
Baiesi-Maes-Wynants formula by the path integral formalism has not been
presented.

Here we consider a general nonequilibrium system, of which dynamics
follows the Langevin equation.
We denote the dynamic variables which obey the Langevin equation as
$X_{1}, X_{2}, \dots, X_{n}$ (with $n$ being the number of
independent stochastic variables).
For convenience, we use the Ito stochastic calculus \cite{Gardiner-book}
for the stochastic
differential equation. (One can employ the Stratonovich
calculus instead of the Ito calculus. Although the calculation below becomes somehow complicated, the
result is essentially the same.)
For simplicity we introduce the shorthand notation $\bm{X} \equiv
(X_{1},X_{2},\dots,X_{n})$.
Because many physical quantities depend on $\bm{X}(t)$, we also introduce
a shorthand notation for a function of $\bm{X}(t)$ as $\hat{f}(t) \equiv
f(\bm{X}(t))$.
The Langevin equation for $\bm{X}(t)$ is described as
\begin{equation}
 \label{lagevin_equation_x}
 \frac{d\bm{X}(t)}{dt} =
 \hat{\bm{V}}(t,h(t))
  + k_{B} T \frac{\partial}{\partial \bm{X}(t)} \cdot \hat{\bm{\Lambda}}(t)
  + \hat{\bm{\xi}}(t) 
\end{equation}
where $\hat{\bm{V}}(t,h(t))$ is the average change rate of $\bm{X}$
(which may be understood as a sort of velocity), $h(t)$ is the external perturbation,
$\hat{\bm{\Lambda}}$ is a positive definite symmetric
tensor, and $\hat{\bm{\xi}}(t)$ is the Gaussian noise.
$\hat{\bm{V}}(t,h(t))$ is decomposed into the reference part which is
independent of $h(t)$ and the perturbation part which is linear in
$h(t)$.
\begin{equation}
 \hat{\bm{V}}(t,h(t)) = \hat{\bm{V}}_{0}(t) + \hat{\bm{V}}_{1}(t) h(t)
\end{equation}
where $\hat{\bm{V}}_{0}(t)$ is the average velocity exerted by the
interaction potential or the external force, and $\hat{\bm{V}}_{1}(t) h(t)$ is
the perturbation term. We assume that $h(t)$ is sufficiently small to
neglect higher order terms in $h(t)$.
The third term in the right hand side of eq \eqref{lagevin_equation_x}
is the stochastic drift term which cancels unphysical probability
current \cite{Risken-book}. $\hat{\bm{\xi}}(t)$ satisfies the following equations.
\begin{align}
 & \label{fluctuation_dissipation_relation_1st_moment_general}
 \langle \hat{\bm{\xi}}(t) \rangle = 0 \\
 & \label{fluctuation_dissipation_relation_2nd_moment_general}
 \langle \hat{\bm{\xi}}(t) \hat{\bm{\xi}}(t') \rangle = 2 k_{B} T
 \hat{\bm{\Lambda}}(t) \delta(t - t')
\end{align}
Eqs \eqref{fluctuation_dissipation_relation_1st_moment_general} and
\eqref{fluctuation_dissipation_relation_2nd_moment_general} can be
interpreted as the fluctuation dissipation type relation.
Or, inversely we can define $\hat{\bm{\Lambda}}(t)$ via eq \eqref{fluctuation_dissipation_relation_2nd_moment_general}.


The probability that a trajectory $\bm{X}(t)$ is realized, which we may call
the path probability (or the path weight), can be calculated
from the distribution of the noise $\hat{\bm{\xi}}(t)$. Because
$\hat{\bm{\xi}}(t)$ obeys the Gaussian distribution, the path
probability $\mathcal{P}[\bm{X}(\cdot)]$ can be calculated as follows \cite{Kleinert-book}.
\begin{equation}
 \label{path_probability_with_perturbation}
  \begin{split}
  \mathcal{P}[\bm{X}(\cdot)] \mathcal{D}\bm{X}
   & = \mathcal{N} \exp\bigg[
   - \frac{1}{4 k_{B} T} \int dt \,
   \left[ \frac{d\bm{X}(t)}{dt}
  - \hat{\bm{V}}(t,h(t))
  - k_{B} T \frac{\partial}{\partial \bm{X}(t)} \cdot \hat{\bm{\Lambda}}(t)
   \right] \\
   & \qquad \cdot \hat{\bm{\Lambda}}^{-1}(t) \cdot
      \left[ \frac{d\bm{X}(t)}{dt}
   - \hat{\bm{V}}(t,h(t))
  - k_{B} T \frac{\partial}{\partial \bm{X}(t)} \cdot \hat{\bm{\Lambda}}(t) \right]
   \bigg] \mathcal{D}\bm{X} \\
   & = \bigg[ 1
   + \frac{1}{2 k_{B} T} \int dt \, 
   \hat{\bm{V}}_{1}(t) \cdot \hat{\bm{\Lambda}}^{-1}(t) \\
   & \qquad   \cdot \left[ \frac{d\bm{X}(t)}{dt}
   - \hat{\bm{V}}_{0}(t)
  - k_{B} T \frac{\partial}{\partial \bm{X}(t)} \cdot
   \hat{\bm{\Lambda}}(t) \right]  h(t)  + O(h^{2})
   \bigg] \mathcal{P}_{0}[\bm{X}(\cdot)] \mathcal{D}\bm{X}
  \end{split}
\end{equation}
where $\mathcal{N}$ is the normalization factor and $\mathcal{P}_{0}[\bm{X}(\cdot)]$ is the path probability at the
reference state (without any perturbations). The time derivative is
interpreted as the retarded derivative \cite{Kleinert-book} (which is
consistent with the Ito calculus), and thus
$\mathcal{N}$ is independent of $h(t)$. The path probability at the
reference is defined as
\begin{equation}
 \label{path_probability_reference}
  \begin{split}
  \mathcal{P}_{0}[\bm{X}(\cdot)] \mathcal{D}\bm{X}
   & \equiv \mathcal{N} \exp\bigg[
   - \frac{1}{4 k_{B} T} \int dt \,
   \left[ \frac{d\bm{X}(t)}{dt}
  - \hat{\bm{V}}_{0}(t)
  - k_{B} T \frac{\partial}{\partial \bm{X}(t)} \cdot \hat{\bm{\Lambda}}(t)
   \right] \\
   & \qquad \cdot \hat{\bm{\Lambda}}^{-1}(t) \cdot
      \left[ \frac{d\bm{X}(t)}{dt}
   - \hat{\bm{V}}_{0}(t)
  - k_{B} T \frac{\partial}{\partial \bm{X}(t)} \cdot \hat{\bm{\Lambda}}(t) \right]
   \bigg] \mathcal{D}\bm{X}
  \end{split}
\end{equation}
By using the path probability \eqref{path_probability_reference}, we can
define the steady state statistical average as the following path integral.
\begin{equation}
 \label{steady_state_statistical_average_path_integral_definition}
  \langle \dotsb \rangle_{\text{ss}}
  \equiv \int \mathcal{D}\bm{X} \dotsb \mathcal{P}_{0}[\bm{X}(\cdot)]
\end{equation}

Since we are interested in the linear response, the $O(h^{2})$ term in
eq \eqref{path_probability_with_perturbation} can be safely
neglected. Then, the statistical average of a physical quantity $A$ at time
$t$ with perturbation can be expressed as
\begin{equation}
 \label{physical_quantity_a_with_perturbation}
 \begin{split}
  \langle \hat{A}(t) \rangle
  & \equiv \int \mathcal{D}\bm{X} \, \hat{A}(t) \mathcal{P}[\bm{X}(\cdot)] \\
  & = A_{\text{ss}}
   + \frac{1}{2 k_{B} T} \int_{-\infty}^{t} dt' \,
  \bigg\langle \hat{A}(t) \hat{\bm{V}}_{1}(t') \cdot \hat{\bm{\Lambda}}^{-1}(t')
  \cdot  \left[ \frac{d\bm{X}(t')}{dt'}
   - \hat{\bm{V}}_{0}(t')
  - k_{B} T \frac{\partial}{\partial \bm{X}(t')} \cdot
  \hat{\bm{\Lambda}}(t') \right] \bigg\rangle_{\text{ss}} h(t')
 \end{split}
\end{equation}
Here, $A_{\text{ss}} \equiv \langle \hat{A}(t) \rangle_{\text{ss}}$ is the steady state statistical average at the
reference state (without perturbation). From time translational
symmetry, $A_{\text{ss}}$ is independent of time $t$.
We have the
following expression as the response function of $A(t)$ to the
perturbation $h(t')$
from eq \eqref{physical_quantity_a_with_perturbation}.
\begin{equation}
 \label{linear_response_formula_path_integral_general}
  \mathcal{R}_{A}(t - t')
  = \frac{1}{2 k_{B} T}
  \bigg\langle \hat{A}(t) \hat{\bm{V}}_{1}(t') \cdot \hat{\bm{\Lambda}}^{-1}(t')
  \cdot  \left[ \frac{d\bm{X}(t')}{dt'}
   - \hat{\bm{V}}_{0}(t')
  - k_{B} T \frac{\partial}{\partial \bm{X}(t')} \cdot
  \hat{\bm{\Lambda}}(t') \right] \bigg\rangle_{\text{ss}}
\end{equation}

Especially, if the perturbation is caused by a perturbation potential,
the perturbation term can be rewritten by using just a single scalar quantity.
If we assume the fluctuation-dissipation type relation, then
$\hat{\bm{V}}_{1}(t)$ can be rewritten as follows.
\begin{equation}
 \hat{\bm{V}}_{1}(t) = \hat{\bm{\Lambda}}(t) \cdot \frac{\partial
  \hat{B}(t)}{\partial \bm{X}(t)}
\end{equation}
Where $B$ is the scalar quantity which is conjugate to $h$.
Eq \eqref{linear_response_formula_path_integral_general} can be
simplified as follows.
\begin{equation}
\label{linear_response_formula_path_integral_potential}
 \begin{split}
  \mathcal{R}_{AB}(t - t')
  & = \frac{1}{2 k_{B} T} \bigg\langle \hat{A}(t)
   \bigg[ \frac{d \hat{B}(t')}{dt'} - \mathcal{L}^{\dagger}
  \hat{B}(t')  \bigg] \bigg\rangle_{\text{ss}} \\
  & = \frac{1}{2 k_{B} T} \frac{d}{dt'} \langle \hat{A}(t)
  \hat{B}(t') \rangle_{\text{ss}}
  - \frac{1}{2 k_{B} T} \langle \hat{A}(t) \mathcal{L}^{\dagger}
  \hat{B}(t') \rangle_{\text{ss}}
 \end{split}
\end{equation}
where we have utilized the Ito formula
\begin{equation}
 \frac{d\hat{B}(t)}{dt}
  = \frac{d\bm{X}(t)}{dt}
  \cdot \frac{\partial \hat{B}(t)}{\partial \bm{X}(t)}
  + k_{B} T \hat{\bm{\Lambda}}(t) : \frac{\partial^{2}
  \hat{B}(t)}{\partial \bm{X}(t) \partial \bm{X}(t)}
\end{equation}
and defined the backward generator $\mathcal{L}^{\dagger}$ (which has the same form as the associate
Fokker-Planck operator) as follows.
\begin{equation}
 \mathcal{L}^{\dagger} \hat{B}(t)
  \equiv 
  \hat{\bm{V}}_{0}(t) \cdot \frac{\partial \hat{B}(t)
  }{\partial \bm{X}(t)}
  + k_{B} T \frac{\partial}{\partial \bm{X}(t)} \cdot \bigg[ \hat{\bm{\Lambda}}(t') \cdot \frac{\partial \hat{B}(t)}{\partial
  \bm{X}(t)} \bigg]
\end{equation}
Eq \eqref{linear_response_formula_path_integral_potential} is nothing
but the Baiesi-Maes-Wynants formula \cite{Baiesi-Maes-Wynants-2009,Baiesi-Maes-Wynants-2009a}. Although eq
\eqref{linear_response_formula_path_integral_potential} is simpler than
eq \eqref{linear_response_formula_path_integral_general},
we should notice that eq
\eqref{linear_response_formula_path_integral_potential} can
be utilized only when the perturbation is given as a perturbation
potential and the fluctuation-dissipation type relation holds.
We should directly use eq
\eqref{linear_response_formula_path_integral_general} if these
conditions are not satisfied.



\section{Weak Kinetic Coupling Between Different Directions}
\label{weak_kinetic_coupling_between_different_directions}

In this appendix, we consider the kinetic coupling effect between the
$x$- and $y$-direction dynamics. From the NEMD simulation results
\cite{Sarman-Evans-Baranyai-1992,Hunt-Todd-2009}, we expect that
the $xy$-element of the mobility tensor is sufficiently small compared
with the diagonal elements.
We can employ the following nondiagonal mobility tensor model as a
simple kinetic coupling model.
\begin{equation}
 \label{mobility_tensor_model_with_nondiagonal_coupling}
  \bm{\Lambda}(\dot{\gamma}) = \frac{1}{\zeta_{0}}
  \begin{bmatrix}
   1 & a \tau_{0} \dot{\gamma} & 0 \\
   a \tau_{0} \dot{\gamma} & 1 & 0 \\
   0 & 0 & 1
  \end{bmatrix}
\end{equation}
Here $a \ll 1$ is a parameter which represents the coupling
strength.
(Eq \eqref{mobility_tensor_model_with_nondiagonal_coupling} is
obtained by setting $\tilde{L}_{2} = \tilde{L}_{3}
= \tilde{L}_{4} = 0$ in eq \eqref{mobility_tensor_general_expansion}.)
Unlike the model considered in the main text, this mobility
model does not accelerate the dynamics in $x$-direction explicitly.
Nonetheless, this model can be used to demonstrate how the kinetic
coupling affects the rheo-dielectric behavior.

We consider $a$ as the perturbation parameter, and expand physical
quantities into the power series of $a$. To see how the kinetic coupling
affects the viscoelastic or dielectric properties, it is sufficient to
consider only the leading order terms (which are proportional to $a$).
First we consider the steady state probability distribution.
Since the Fokker-Planck
equation is linear, the steady state probability distribution can
be expressed as a Gaussian. Then the covariance matrix can be
expressed as follows.
\begin{equation}
 \label{steady_state_covariance_matrix_with_nondiagonal_coupling}
  \bm{C}(\dot{\gamma}) =
  \frac{\bar{R}^{2}}{3} 
  \begin{bmatrix}
   \displaystyle 1
  + \frac{(\tau_{0} \dot{\gamma})^{2}}{2} &
  \displaystyle \frac{\tau_{0} \dot{\gamma}}{2} & 0 \\
  \displaystyle \frac{\tau_{0} \dot{\gamma}}{2} & 1 & 0 \\
  0 & 0 & 1
  \end{bmatrix}
  - \frac{\bar{R}^{2}}{3} \frac{a (\tau_{0}
  \dot{\gamma})^{2}}{2}
  \begin{bmatrix}
   1 + (\tau_{0} \dot{\gamma})^{2}
   & \tau_{0} \dot{\gamma}
   & 0 \\
   \tau_{0} \dot{\gamma} &
   1 & 0 \\
  0 & 0 & 0
  \end{bmatrix}
  + O(a^{2})
\end{equation}

\begin{equation}
 \begin{split}
 \langle R_{y}(t) R_{y}(0) \rangle_{\text{ss}}
  & = C_{yy} e^{- t / \tau_{0}}
  - a \dot{\gamma} \int_{0}^{t} dt' \, e^{- (t - t') / \tau_{0}} \langle R_{x}(t') R_{y}(0)
  \rangle_{\text{ss}} \\
  & = C_{yy} e^{- t / \tau_{0}}
  - a 
  \left[ C_{xy} \dot{\gamma} t e^{- t / \tau_{0}}
  + \frac{1}{2} C_{yy} \dot{\gamma}^{2} t^{2} e^{- t / \tau_{0}}
  \right] + O(a^{2}) \\
 \end{split}
\end{equation}
\begin{equation}
 \langle R_{y}(t) R_{x}(0) \rangle_{\text{ss}}
  = C_{xy} e^{- t / \tau_{0}} + O(a)
\end{equation}

\begin{figure}[h!]
\includegraphics{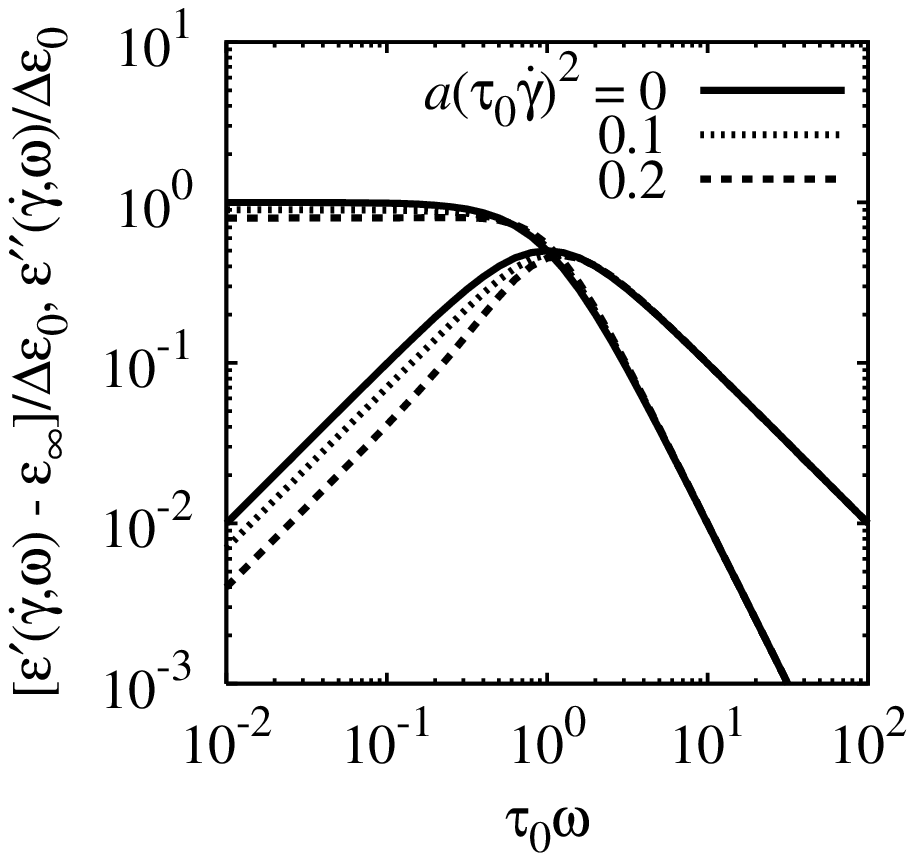}
\caption{\label{rheo_dielectric_weak_coupling}}
\end{figure}

From the Baiesi-Maes-Wynants formula, the rheo-dielectric response
function becomes as follows.
\begin{equation}
 \label{rheo_dielectric_response_function_weak_coupling}
 \begin{split}
 \varphi(\dot{\gamma},t)
  & =
  \frac{3 \Delta\varepsilon_{0}}{2 \bar{R}^{2}} \bigg[ - \frac{d}{dt} \langle R_{y}(t) R_{y}(0)
   \rangle_{\text{ss}}
   + \frac{1}{\tau_{0}} \langle R_{y}(t)
   [R_{y}(0) + a \tau_{0} \dot{\gamma} R_{x}(0)] \rangle_{\text{ss}}
   \bigg] \\
  & =
  \Delta\varepsilon_{0} \frac{1}{\tau_{0}} e^{- t / \tau_{0}} \left[
  1 - \frac{1}{2} a \dot{\gamma}^{2} t^{2}
  + O(a^{2})
   \right]
 \end{split}
\end{equation}
The rheo-dielectric intensity can be calculated to be
\begin{equation}
 \label{rheo_dielectric_intensity_weak_coupling}
 \Delta\varepsilon(\dot{\gamma})
  = \int_{0}^{\infty} dt \, \varphi(t,\dot{\gamma})
  = \Delta\varepsilon_{0}
  [ 1 - a (\tau_{0}\dot{\gamma})^{2}  + O(a^{2}) ]
\end{equation}
From eq \eqref{rheo_dielectric_intensity_weak_coupling} we find that the
rheo-dielectric intensity is decreased by the kinetic coupling. This is
in contrast to the case of the anisotropic mobility model,
where the rheo-dielectric intensity is independent of the shear rate.
By performing the Fourier transform for eq
\eqref{rheo_dielectric_response_function_weak_coupling}, we have the
following real and imaginary parts of the rheo-dielectric response function in the frequency domain.
\begin{align}
 & \label{epsilon_prime_weak_coupling}
 \varepsilon'(\dot{\gamma},\omega)
  - \varepsilon_{\infty} 
  = \Delta\varepsilon_{0}
  \bigg[
  \frac{1}{1 + (\tau_{0} \omega)^{2}}
  - a (\tau_{0} \dot{\gamma})^{2}
  \frac{1 - 3 (\tau_{0} \omega)^{2}}{[1 + (\tau_{0} \omega)^{2}]^{3}}
  + O(a^{2})
   \bigg] \\
 & \label{epsilon_double_prime_weak_coupling}
 \varepsilon''(\dot{\gamma},\omega)
  = \Delta\varepsilon_{0}
  \bigg[
  \frac{\tau_{0} \omega}{1 + (\tau_{0} \omega)^{2}}
  - a (\tau_{0} \dot{\gamma})^{2}
  \frac{3 \tau_{0} \omega  - (\tau_{0} \omega)^{3}}{[1 + (\tau_{0} \omega)^{2}]^{3}}
  + O(a^{2})
   \bigg]
\end{align}
From eqs \eqref{epsilon_prime_weak_coupling} and
\eqref{epsilon_double_prime_weak_coupling}, we find that
the effects of the kinetic
coupling and the shear rate can be represented by an
effective coupling constant, $a (\tau_{0} \dot{\gamma})^{2}$.
We show $\epsilon'(\dot{\gamma},\omega)$ and
$\epsilon''(\dot{\gamma},\omega)$ by eqs \eqref{epsilon_prime_weak_coupling} and
\eqref{epsilon_double_prime_weak_coupling}
in FIG. \ref{rheo_dielectric_weak_coupling}.
We can observe that the dielectric loss
$\epsilon''(\dot{\gamma},\omega)$ for $\tau_{0} \omega \lesssim 1$
decreases as the effective coupling constant $a (\tau_{0} \dot{\gamma})^{2}$
increases. On the other hand, we can observe that the rheo-dielectric
response functions are insensitive to the shear rate for $\tau_{0} \omega
\gtrsim 1$. These are qualitatively similar to experimentally observed
rheo-dielectric behavior of entangled star polymers \cite{Watanabe-Ishida-Matsumiya-2002}.

\section{Anisotropic Version of the Convective Constraint Release Model}
\label{anisotropic_version_of_the_convective_constraint_release_model}

As we discussed in the main text, the convective constraint release
(CCR) model isotropically accelerates the dynamics of a polymer chain.
In order to reproduce the rheo-dielectric response behavior, we need to
modify the CCR model to be anisotropic. In this appendix, we consider a
simple anisotropic version of the CCR model, as a possible modification
for the CCR model.

The key feature of the CCR model is that the characteristic relaxation
time is accelerated by the product of the velocity gradient tensor and
the chain conformation tensor. To make the CCR model anisotropic, here we
employ the following mobility tensor.
\begin{equation}
 \label{mobility_tensor_anisotropic_ccr}
 \bm{\Lambda}_{\text{CCR}}(\dot{\gamma})
  = \frac{1}{\zeta_{0}}  \bm{1} 
     + \frac{\beta_{\text{CCR}}'}{2 k_{B} T}
     \big[ \bm{\kappa} \cdot (\langle \bm{R} \bm{R}
   \rangle - \langle \bm{R} \bm{R} \rangle_{\text{eq}})
     + (\langle \bm{R} \bm{R}
   \rangle - \langle \bm{R} \bm{R} \rangle_{\text{eq}}) \cdot \bm{\kappa}^{t} \big]
\end{equation}
where $\beta_{\text{CCR}}'$ is a positive constant of the order of unity
and the velocity gradient tensor $\bm{\kappa}$ is given by eq
\eqref{velocity_gradient_tensor_simple_shear}.
Eq \eqref{mobility_tensor_anisotropic_ccr} is similar to eq
\eqref{mobility_tensor_ccr} but the scalar diadic product $\bm{\kappa} :
\langle \bm{R} \bm{R} \rangle$ is replaced by second rank tensor products.
For the steady state under shear, the ensemble average $\langle \bm{R} \bm{R}
\rangle$ is replaced by the steady state average $\langle \bm{R} \bm{R}
\rangle_{\text{ss}}$, and we can obtain the steady state probability
distribution explicitly. The anisotropic
CCR model \eqref{mobility_tensor_anisotropic_ccr} has the Gaussian form
steady state probability distribution
\eqref{steady_state_probability_distribution} with the following
covariance matrix.
\begin{equation}
 \label{steady_state_covariance_matrix_anisotropic_ccr}
 \bm{C}(\dot{\gamma}) \equiv
  \frac{\bar{R}^{2}}{3}
  \begin{bmatrix}
   \displaystyle 1 + \frac{\sqrt{\beta_{\text{CCR}}'
  (\tau_{0} \dot{\gamma})^{2} + 1} - 1}{\beta_{\text{CCR}}' \sqrt{\beta_{\text{CCR}}'
  (\tau_{0} \dot{\gamma})^{2} + 1}} &
  \displaystyle \frac{\sqrt{\beta_{\text{CCR}}'
  (\tau_{0} \dot{\gamma})^{2} + 1} - 1}{\beta_{\text{CCR}}' \tau_{0}
   \dot{\gamma}} & 0 \\[1.2em]
  \displaystyle \frac{\sqrt{\beta_{\text{CCR}}'
  (\tau_{0} \dot{\gamma})^{2} + 1} - 1}{\beta_{\text{CCR}}' \tau_{0} \dot{\gamma}} & 1 & 0 \\[1.2em]
  0 & 0 & 1
  \end{bmatrix}
\end{equation}
It is straightforward to show that eqs
\eqref{steady_state_probability_distribution},
\eqref{mobility_tensor_anisotropic_ccr}, and
\eqref{steady_state_covariance_matrix_anisotropic_ccr} satisfy the
steady state condition \eqref{steady_state_probability_distribution_condition}.
The steady state viscosity is expressed as follows.
\begin{equation}
\label{steady_state_viscosity_anisotropic_ccr}
 \eta(\dot{\gamma}) = \eta_{0} \frac{2 \big[ \sqrt{\beta_{\text{CCR}}'
  (\tau_{0} \dot{\gamma})^{2} + 1} - 1 \big]}{\beta_{\text{CCR}}' (\tau_{0} \dot{\gamma})^{2}}
\end{equation}
Eq \eqref{steady_state_viscosity_anisotropic_ccr} is a monotonically
decreasing function of $\dot{\gamma}^{2}$, and thus the anisotropic CCR
model shows the shear thinning behavior. Eq
\eqref{steady_state_viscosity_anisotropic_ccr} is similar to that for the steady state
shear viscosity in the original CCR model.
At the high shear rate region, eq
\eqref{steady_state_viscosity_anisotropic_ccr} approaches
to the following asymptotic form.
\begin{equation}
 \label{steady_state_viscosity_anisotropic_ccr_asymptotic}
  \eta(\dot{\gamma}) \to \eta_{0} \frac{2}{\sqrt{\beta_{\text{CCR}}'} \tau_{0}
  \dot{\gamma}} \qquad (\dot{\gamma} \to \infty)
\end{equation}
As in the original CCR model \cite{Marrucci-1996}, eq
\eqref{steady_state_viscosity_anisotropic_ccr_asymptotic} becomes consistent with
the Cox-Merz rule when we set $\beta_{\text{CCR}}' = 4$. We can
straightforwardly show that the first normal stress
difference coefficient by the anisotropic CCR model is also similar to
one by the original CCR model.

The steady state covariance matrix
\eqref{steady_state_covariance_matrix_anisotropic_ccr} looks similar to
the steady state covariance matrix for the anisotropic mobility model \eqref{steady_state_covariance_matrix}.
In fact, the anisotropic CCR model reduces to the anisotropic mobility tensor
model \eqref{mobility_tensor_model}.
Substituting eq \eqref{steady_state_covariance_matrix_anisotropic_ccr}
into eq \eqref{mobility_tensor_anisotropic_ccr}, we have
\begin{equation}
 \label{mobility_tensor_anisotropic_ccr_explicit}
 \bm{\Lambda}_{\text{CCR}}(\dot{\gamma})
  = \frac{1}{\zeta_{0}}
  \begin{bmatrix}
   \sqrt{\beta_{\text{CCR}}' (\tau_{0} \dot{\gamma})^{2} + 1} & 0 & 0 \\
   0 & 1 & 0 \\
   0 & 0 & 1
  \end{bmatrix}
\end{equation}
This has the same form as the power law type anisotropic mobility tensor model
(eqs \eqref{mobility_tensor_model} and \eqref{dimensionless_mobility_factor_power_law}) with $\tau_{c} =
\sqrt{\beta_{\text{CCR}}'} \tau_{0}$ and $\alpha = 1$.
Then it is clear that the anisotropic CCR model successfully reproduces
essential properties of our anisotropic mobility tensor model.
However, it should be noticed that the anisotropic CCR mobility
tensor \eqref{mobility_tensor_anisotropic_ccr} depends
on $\langle \bm{R} \bm{R} \rangle$ in the presence of the external
perturbation field.
Because $\langle \bm{R} \bm{R} \rangle$
generally depends on the applied perturbation field implicitly, the
anisotropic CCR mobility tensor  gives
additional contributions for linear response functions.
Thus the linear response properties of the anisotropic CCR model will be
slightly different from ones of our anisotropic mobility model.
We should carefully calculate the contribution from the mobility tensor
to get the linear response functions for the anisotropic CCR model.
(Fortunately, even if the mobility tensor depends on the perturbation field,
the path integral formulation shown in Appendix
\ref{derivation_of_the_baiesi_maes_wynants_formula_by_the_path_integral_formalism}
is still valid.)
\bibliographystyle{apsrev4-1}
\bibliography{anisotropic_mobility_model}

\begin{thebibliography}{86}%
\makeatletter
\providecommand \@ifxundefined [1]{%
 \@ifx{#1\undefined}
}%
\providecommand \@ifnum [1]{%
 \ifnum #1\expandafter \@firstoftwo
 \else \expandafter \@secondoftwo
 \fi
}%
\providecommand \@ifx [1]{%
 \ifx #1\expandafter \@firstoftwo
 \else \expandafter \@secondoftwo
 \fi
}%
\providecommand \natexlab [1]{#1}%
\providecommand \enquote  [1]{``#1''}%
\providecommand \bibnamefont  [1]{#1}%
\providecommand \bibfnamefont [1]{#1}%
\providecommand \citenamefont [1]{#1}%
\providecommand \href@noop [0]{\@secondoftwo}%
\providecommand \href [0]{\begingroup \@sanitize@url \@href}%
\providecommand \@href[1]{\@@startlink{#1}\@@href}%
\providecommand \@@href[1]{\endgroup#1\@@endlink}%
\providecommand \@sanitize@url [0]{\catcode `\\12\catcode `\$12\catcode
  `\&12\catcode `\#12\catcode `\^12\catcode `\_12\catcode `\%12\relax}%
\providecommand \@@startlink[1]{}%
\providecommand \@@endlink[0]{}%
\providecommand \url  [0]{\begingroup\@sanitize@url \@url }%
\providecommand \@url [1]{\endgroup\@href {#1}{\urlprefix }}%
\providecommand \urlprefix  [0]{URL }%
\providecommand \Eprint [0]{\href }%
\providecommand \doibase [0]{http://dx.doi.org/}%
\providecommand \selectlanguage [0]{\@gobble}%
\providecommand \bibinfo  [0]{\@secondoftwo}%
\providecommand \bibfield  [0]{\@secondoftwo}%
\providecommand \translation [1]{[#1]}%
\providecommand \BibitemOpen [0]{}%
\providecommand \bibitemStop [0]{}%
\providecommand \bibitemNoStop [0]{.\EOS\space}%
\providecommand \EOS [0]{\spacefactor3000\relax}%
\providecommand \BibitemShut  [1]{\csname bibitem#1\endcsname}%
\let\auto@bib@innerbib\@empty
\bibitem [{\citenamefont {Kubo}\ \emph {et~al.}(1991)\citenamefont {Kubo},
  \citenamefont {Toda},\ and\ \citenamefont
  {Hashitsume}}]{Kubo-Toada-Hashitsume-book}%
  \BibitemOpen
  \bibfield  {author} {\bibinfo {author} {\bibfnamefont {R.}~\bibnamefont
  {Kubo}}, \bibinfo {author} {\bibfnamefont {M.}~\bibnamefont {Toda}}, \ and\
  \bibinfo {author} {\bibfnamefont {N.}~\bibnamefont {Hashitsume}},\
  }\href@noop {} {\emph {\bibinfo {title} {Statistical Physics II}}}\ (\bibinfo
   {publisher} {Springer},\ \bibinfo {address} {Heidelberg},\ \bibinfo {year}
  {1991})\BibitemShut {NoStop}%
\bibitem [{\citenamefont {Evans}\ and\ \citenamefont
  {Morris}(2008)}]{Evans-Morris-book}%
  \BibitemOpen
  \bibfield  {author} {\bibinfo {author} {\bibfnamefont {D.~J.}\ \bibnamefont
  {Evans}}\ and\ \bibinfo {author} {\bibfnamefont {G.~P.}\ \bibnamefont
  {Morris}},\ }\href@noop {} {\emph {\bibinfo {title} {Statistical Mechanics of
  Nonequilibrium Liquids}}},\ \bibinfo {edition} {2nd}\ ed.\ (\bibinfo
  {publisher} {Cambridge University Press},\ \bibinfo {address} {Cambridge},\
  \bibinfo {year} {2008})\BibitemShut {NoStop}%
\bibitem [{\citenamefont {Risken}(1989)}]{Risken-book}%
  \BibitemOpen
  \bibfield  {author} {\bibinfo {author} {\bibfnamefont {H.}~\bibnamefont
  {Risken}},\ }\href@noop {} {\emph {\bibinfo {title} {The Fokker-Planck
  Equation}}},\ \bibinfo {edition} {2nd}\ ed.\ (\bibinfo  {publisher}
  {Springer},\ \bibinfo {address} {Berlin},\ \bibinfo {year}
  {1989})\BibitemShut {NoStop}%
\bibitem [{\citenamefont {Doi}\ and\ \citenamefont
  {Edwards}(1986)}]{Doi-Edwards-book}%
  \BibitemOpen
  \bibfield  {author} {\bibinfo {author} {\bibfnamefont {M.}~\bibnamefont
  {Doi}}\ and\ \bibinfo {author} {\bibfnamefont {S.~F.}\ \bibnamefont
  {Edwards}},\ }\href@noop {} {\emph {\bibinfo {title} {The Theory of Polymer
  Dynamics}}}\ (\bibinfo  {publisher} {Oxford University Press},\ \bibinfo
  {address} {Oxford},\ \bibinfo {year} {1986})\BibitemShut {NoStop}%
\bibitem [{\citenamefont {Watanabe}(1999)}]{Watanabe-1999}%
  \BibitemOpen
  \bibfield  {author} {\bibinfo {author} {\bibfnamefont {H.}~\bibnamefont
  {Watanabe}},\ }\href@noop {} {\bibfield  {journal} {\bibinfo  {journal}
  {Prog. Polym. Sci.}\ }\textbf {\bibinfo {volume} {24}},\ \bibinfo {pages}
  {1253} (\bibinfo {year} {1999})}\BibitemShut {NoStop}%
\bibitem [{\citenamefont {Matsumiya}\ \emph {et~al.}(2000)\citenamefont
  {Matsumiya}, \citenamefont {Watanabe},\ and\ \citenamefont
  {Osaki}}]{Matsumiya-Watanabe-Osaki-2000}%
  \BibitemOpen
  \bibfield  {author} {\bibinfo {author} {\bibfnamefont {Y.}~\bibnamefont
  {Matsumiya}}, \bibinfo {author} {\bibfnamefont {H.}~\bibnamefont {Watanabe}},
  \ and\ \bibinfo {author} {\bibfnamefont {K.}~\bibnamefont {Osaki}},\
  }\href@noop {} {\bibfield  {journal} {\bibinfo  {journal} {Macromolecules}\
  }\textbf {\bibinfo {volume} {33}},\ \bibinfo {pages} {499} (\bibinfo {year}
  {2000})}\BibitemShut {NoStop}%
\bibitem [{\citenamefont {Watanabe}\ \emph {et~al.}(2000)\citenamefont
  {Watanabe}, \citenamefont {Matsumiya},\ and\ \citenamefont
  {Osaki}}]{Watanabe-Matsumiya-Osaki-2000}%
  \BibitemOpen
  \bibfield  {author} {\bibinfo {author} {\bibfnamefont {H.}~\bibnamefont
  {Watanabe}}, \bibinfo {author} {\bibfnamefont {Y.}~\bibnamefont {Matsumiya}},
  \ and\ \bibinfo {author} {\bibfnamefont {K.}~\bibnamefont {Osaki}},\
  }\href@noop {} {\bibfield  {journal} {\bibinfo  {journal} {J. Polym. Sci. B:
  Polym. Phys.}\ }\textbf {\bibinfo {volume} {38}},\ \bibinfo {pages} {1024}
  (\bibinfo {year} {2000})}\BibitemShut {NoStop}%
\bibitem [{\citenamefont {Matsumiya}\ and\ \citenamefont
  {Watanabe}(2001)}]{Matsumiya-Watanabe-2001}%
  \BibitemOpen
  \bibfield  {author} {\bibinfo {author} {\bibfnamefont {Y.}~\bibnamefont
  {Matsumiya}}\ and\ \bibinfo {author} {\bibfnamefont {H.}~\bibnamefont
  {Watanabe}},\ }\href@noop {} {\bibfield  {journal} {\bibinfo  {journal}
  {Macromolecules}\ }\textbf {\bibinfo {volume} {34}},\ \bibinfo {pages} {5702}
  (\bibinfo {year} {2001})}\BibitemShut {NoStop}%
\bibitem [{\citenamefont {Watanabe}\ \emph
  {et~al.}(2002{\natexlab{a}})\citenamefont {Watanabe}, \citenamefont
  {Matsumiya},\ and\ \citenamefont {Inoue}}]{Watanabe-Matsumiya-Inoue-2002}%
  \BibitemOpen
  \bibfield  {author} {\bibinfo {author} {\bibfnamefont {H.}~\bibnamefont
  {Watanabe}}, \bibinfo {author} {\bibfnamefont {Y.}~\bibnamefont {Matsumiya}},
  \ and\ \bibinfo {author} {\bibfnamefont {T.}~\bibnamefont {Inoue}},\
  }\href@noop {} {\bibfield  {journal} {\bibinfo  {journal} {Macromolecules}\
  }\textbf {\bibinfo {volume} {35}},\ \bibinfo {pages} {2339} (\bibinfo {year}
  {2002}{\natexlab{a}})}\BibitemShut {NoStop}%
\bibitem [{\citenamefont {Osaki}\ \emph {et~al.}(1965)\citenamefont {Osaki},
  \citenamefont {Tamura}, \citenamefont {Kurata},\ and\ \citenamefont
  {Kotaka}}]{Osaki-Tamura-Kurata-Kotaka-1965}%
  \BibitemOpen
  \bibfield  {author} {\bibinfo {author} {\bibfnamefont {K.}~\bibnamefont
  {Osaki}}, \bibinfo {author} {\bibfnamefont {M.}~\bibnamefont {Tamura}},
  \bibinfo {author} {\bibfnamefont {M.}~\bibnamefont {Kurata}}, \ and\ \bibinfo
  {author} {\bibfnamefont {T.}~\bibnamefont {Kotaka}},\ }\href@noop {}
  {\bibfield  {journal} {\bibinfo  {journal} {J. Phys. Chem.}\ }\textbf
  {\bibinfo {volume} {69}},\ \bibinfo {pages} {4183} (\bibinfo {year}
  {1965})}\BibitemShut {NoStop}%
\bibitem [{\citenamefont {de~L.~Costello}(1997)}]{Costello-1977}%
  \BibitemOpen
  \bibfield  {author} {\bibinfo {author} {\bibfnamefont {B.~A.}\ \bibnamefont
  {de~L.~Costello}},\ }\href@noop {} {\bibfield  {journal} {\bibinfo  {journal}
  {J. Non-Newtonian Fluid Mech.}\ }\textbf {\bibinfo {volume} {68}},\ \bibinfo
  {pages} {303} (\bibinfo {year} {1997})}\BibitemShut {NoStop}%
\bibitem [{\citenamefont {Vermant}\ \emph {et~al.}(1998)\citenamefont
  {Vermant}, \citenamefont {Walker}, \citenamefont {Moldenears},\ and\
  \citenamefont {Mewis}}]{Vermant-Walker-Moldenaers-Mewis-1998}%
  \BibitemOpen
  \bibfield  {author} {\bibinfo {author} {\bibfnamefont {J.}~\bibnamefont
  {Vermant}}, \bibinfo {author} {\bibfnamefont {L.}~\bibnamefont {Walker}},
  \bibinfo {author} {\bibfnamefont {P.}~\bibnamefont {Moldenears}}, \ and\
  \bibinfo {author} {\bibfnamefont {J.}~\bibnamefont {Mewis}},\ }\href@noop {}
  {\bibfield  {journal} {\bibinfo  {journal} {J. Non-Newtonian Fluid Mech.}\
  }\textbf {\bibinfo {volume} {79}},\ \bibinfo {pages} {173} (\bibinfo {year}
  {1998})}\BibitemShut {NoStop}%
\bibitem [{\citenamefont {Isayev}\ and\ \citenamefont
  {Wong}(1988)}]{Isayev-Wong-1988}%
  \BibitemOpen
  \bibfield  {author} {\bibinfo {author} {\bibfnamefont {A.~I.}\ \bibnamefont
  {Isayev}}\ and\ \bibinfo {author} {\bibfnamefont {C.~M.}\ \bibnamefont
  {Wong}},\ }\href@noop {} {\bibfield  {journal} {\bibinfo  {journal} {J.
  Polym. Sci. B: Polym. Phys.}\ }\textbf {\bibinfo {volume} {26}},\ \bibinfo
  {pages} {2303} (\bibinfo {year} {1988})}\BibitemShut {NoStop}%
\bibitem [{\citenamefont {Wong}\ and\ \citenamefont
  {Isayev}(1989)}]{Wong-Isayev-1989}%
  \BibitemOpen
  \bibfield  {author} {\bibinfo {author} {\bibfnamefont {C.~M.}\ \bibnamefont
  {Wong}}\ and\ \bibinfo {author} {\bibfnamefont {A.~I.}\ \bibnamefont
  {Isayev}},\ }\href@noop {} {\bibfield  {journal} {\bibinfo  {journal} {Rheol.
  Acta}\ }\textbf {\bibinfo {volume} {28}},\ \bibinfo {pages} {176} (\bibinfo
  {year} {1989})}\BibitemShut {NoStop}%
\bibitem [{\citenamefont {Somma}\ \emph {et~al.}(2007)\citenamefont {Somma},
  \citenamefont {Valentino}, \citenamefont {Titomanlio},\ and\ \citenamefont
  {Ianniruberto}}]{Somma-Valentino-Titomanlio-Ianniruberto-2007}%
  \BibitemOpen
  \bibfield  {author} {\bibinfo {author} {\bibfnamefont {E.}~\bibnamefont
  {Somma}}, \bibinfo {author} {\bibfnamefont {O.}~\bibnamefont {Valentino}},
  \bibinfo {author} {\bibfnamefont {G.}~\bibnamefont {Titomanlio}}, \ and\
  \bibinfo {author} {\bibfnamefont {G.}~\bibnamefont {Ianniruberto}},\
  }\href@noop {} {\bibfield  {journal} {\bibinfo  {journal} {J. Rheol.}\
  }\textbf {\bibinfo {volume} {51}},\ \bibinfo {pages} {987} (\bibinfo {year}
  {2007})}\BibitemShut {NoStop}%
\bibitem [{\citenamefont {Boukany}\ and\ \citenamefont
  {Wang}(2009)}]{Boukany-Wang-2009}%
  \BibitemOpen
  \bibfield  {author} {\bibinfo {author} {\bibfnamefont {P.~E.}\ \bibnamefont
  {Boukany}}\ and\ \bibinfo {author} {\bibfnamefont {S.-Q.}\ \bibnamefont
  {Wang}},\ }\href@noop {} {\bibfield  {journal} {\bibinfo  {journal} {J.
  Rheol.}\ }\textbf {\bibinfo {volume} {53}},\ \bibinfo {pages} {1425}
  (\bibinfo {year} {2009})}\BibitemShut {NoStop}%
\bibitem [{\citenamefont {Li}\ and\ \citenamefont {Wang}(2010)}]{Li-Wang-2010}%
  \BibitemOpen
  \bibfield  {author} {\bibinfo {author} {\bibfnamefont {X.}~\bibnamefont
  {Li}}\ and\ \bibinfo {author} {\bibfnamefont {S.-Q.}\ \bibnamefont {Wang}},\
  }\href@noop {} {\bibfield  {journal} {\bibinfo  {journal} {Macromolecules}\
  }\textbf {\bibinfo {volume} {43}},\ \bibinfo {pages} {5904} (\bibinfo {year}
  {2010})}\BibitemShut {NoStop}%
\bibitem [{\citenamefont {Matsumiya}\ \emph {et~al.}(1998)\citenamefont
  {Matsumiya}, \citenamefont {Watanabe}, \citenamefont {Inoue}, \citenamefont
  {Osaki},\ and\ \citenamefont
  {Yao}}]{Matsumiya-Watanabe-Inoue-Osaki-Yao-1998}%
  \BibitemOpen
  \bibfield  {author} {\bibinfo {author} {\bibfnamefont {Y.}~\bibnamefont
  {Matsumiya}}, \bibinfo {author} {\bibfnamefont {H.}~\bibnamefont {Watanabe}},
  \bibinfo {author} {\bibfnamefont {T.}~\bibnamefont {Inoue}}, \bibinfo
  {author} {\bibfnamefont {K.}~\bibnamefont {Osaki}}, \ and\ \bibinfo {author}
  {\bibfnamefont {M.-L.}\ \bibnamefont {Yao}},\ }\href@noop {} {\bibfield
  {journal} {\bibinfo  {journal} {Macromolecules}\ }\textbf {\bibinfo {volume}
  {31}},\ \bibinfo {pages} {7973} (\bibinfo {year} {1998})}\BibitemShut
  {NoStop}%
\bibitem [{\citenamefont {Watanabe}\ \emph {et~al.}(1999)\citenamefont
  {Watanabe}, \citenamefont {Sato}, \citenamefont {Matsumiya}, \citenamefont
  {Inoue},\ and\ \citenamefont
  {Osaki}}]{Watanabe-Sato-Matsumiya-Inoue-Osaki-1999}%
  \BibitemOpen
  \bibfield  {author} {\bibinfo {author} {\bibfnamefont {H.}~\bibnamefont
  {Watanabe}}, \bibinfo {author} {\bibfnamefont {T.}~\bibnamefont {Sato}},
  \bibinfo {author} {\bibfnamefont {Y.}~\bibnamefont {Matsumiya}}, \bibinfo
  {author} {\bibfnamefont {T.}~\bibnamefont {Inoue}}, \ and\ \bibinfo {author}
  {\bibfnamefont {K.}~\bibnamefont {Osaki}},\ }\href@noop {} {\bibfield
  {journal} {\bibinfo  {journal} {Nihon Reoroji Gakkaishi (J. Soc. Rheol.
  Jpn.)}\ }\textbf {\bibinfo {volume} {27}},\ \bibinfo {pages} {121} (\bibinfo
  {year} {1999})}\BibitemShut {NoStop}%
\bibitem [{\citenamefont {Watanabe}\ \emph
  {et~al.}(2002{\natexlab{b}})\citenamefont {Watanabe}, \citenamefont
  {Ishida},\ and\ \citenamefont {Matsumiya}}]{Watanabe-Ishida-Matsumiya-2002}%
  \BibitemOpen
  \bibfield  {author} {\bibinfo {author} {\bibfnamefont {H.}~\bibnamefont
  {Watanabe}}, \bibinfo {author} {\bibfnamefont {S.}~\bibnamefont {Ishida}}, \
  and\ \bibinfo {author} {\bibfnamefont {Y.}~\bibnamefont {Matsumiya}},\
  }\href@noop {} {\bibfield  {journal} {\bibinfo  {journal} {Macromolecules}\
  }\textbf {\bibinfo {volume} {35}},\ \bibinfo {pages} {8802} (\bibinfo {year}
  {2002}{\natexlab{b}})}\BibitemShut {NoStop}%
\bibitem [{\citenamefont {Watanabe}\ \emph
  {et~al.}(2003{\natexlab{a}})\citenamefont {Watanabe}, \citenamefont
  {Matsumiya},\ and\ \citenamefont {Inoue}}]{Watanabe-Matsumiya-Inoue-2003}%
  \BibitemOpen
  \bibfield  {author} {\bibinfo {author} {\bibfnamefont {H.}~\bibnamefont
  {Watanabe}}, \bibinfo {author} {\bibfnamefont {Y.}~\bibnamefont {Matsumiya}},
  \ and\ \bibinfo {author} {\bibfnamefont {T.}~\bibnamefont {Inoue}},\
  }\href@noop {} {\bibfield  {journal} {\bibinfo  {journal} {J. Phys.: Cond.
  Matt.}\ }\textbf {\bibinfo {volume} {15}},\ \bibinfo {pages} {S909} (\bibinfo
  {year} {2003}{\natexlab{a}})}\BibitemShut {NoStop}%
\bibitem [{\citenamefont {Watanabe}\ \emph {et~al.}(2005)\citenamefont
  {Watanabe}, \citenamefont {Matsumiya},\ and\ \citenamefont
  {Inoue}}]{Watanabe-Matsumiya-Inoue-2005}%
  \BibitemOpen
  \bibfield  {author} {\bibinfo {author} {\bibfnamefont {H.}~\bibnamefont
  {Watanabe}}, \bibinfo {author} {\bibfnamefont {Y.}~\bibnamefont {Matsumiya}},
  \ and\ \bibinfo {author} {\bibfnamefont {T.}~\bibnamefont {Inoue}},\
  }\href@noop {} {\bibfield  {journal} {\bibinfo  {journal} {Macromol. Symp.}\
  }\textbf {\bibinfo {volume} {228}},\ \bibinfo {pages} {51} (\bibinfo {year}
  {2005})}\BibitemShut {NoStop}%
\bibitem [{\citenamefont {Uneyama}\ \emph {et~al.}(2009)\citenamefont
  {Uneyama}, \citenamefont {Masubuchi}, \citenamefont {Horio}, \citenamefont
  {Matsumiya}, \citenamefont {Watanabe}, \citenamefont {Pathak},\ and\
  \citenamefont
  {Roland}}]{Uneyama-Masubuchi-Horio-Matsumiya-Watanabe-Pathak-Roland-2009}%
  \BibitemOpen
  \bibfield  {author} {\bibinfo {author} {\bibfnamefont {T.}~\bibnamefont
  {Uneyama}}, \bibinfo {author} {\bibfnamefont {Y.}~\bibnamefont {Masubuchi}},
  \bibinfo {author} {\bibfnamefont {K.}~\bibnamefont {Horio}}, \bibinfo
  {author} {\bibfnamefont {Y.}~\bibnamefont {Matsumiya}}, \bibinfo {author}
  {\bibfnamefont {H.}~\bibnamefont {Watanabe}}, \bibinfo {author}
  {\bibfnamefont {J.~A.}\ \bibnamefont {Pathak}}, \ and\ \bibinfo {author}
  {\bibfnamefont {C.~M.}\ \bibnamefont {Roland}},\ }\href@noop {} {\bibfield
  {journal} {\bibinfo  {journal} {J. Polym. Sci. B: Polym. Phys.}\ }\textbf
  {\bibinfo {volume} {47}},\ \bibinfo {pages} {1039} (\bibinfo {year}
  {2009})}\BibitemShut {NoStop}%
\bibitem [{\citenamefont {Masubuchi}\ \emph {et~al.}(2004)\citenamefont
  {Masubuchi}, \citenamefont {Watanabe}, \citenamefont {Ianniruberto},
  \citenamefont {Greco},\ and\ \citenamefont
  {Marrucci}}]{Masubuchi-Watanabe-Ianniruberto-Greco-Marrucci-2004}%
  \BibitemOpen
  \bibfield  {author} {\bibinfo {author} {\bibfnamefont {Y.}~\bibnamefont
  {Masubuchi}}, \bibinfo {author} {\bibfnamefont {H.}~\bibnamefont {Watanabe}},
  \bibinfo {author} {\bibfnamefont {G.}~\bibnamefont {Ianniruberto}}, \bibinfo
  {author} {\bibfnamefont {F.}~\bibnamefont {Greco}}, \ and\ \bibinfo {author}
  {\bibfnamefont {G.}~\bibnamefont {Marrucci}},\ }\href@noop {} {\bibfield
  {journal} {\bibinfo  {journal} {Nihon Reoroji Gakkaishi (J. Soc. Rheol.
  Jpn.)}\ }\textbf {\bibinfo {volume} {32}},\ \bibinfo {pages} {192} (\bibinfo
  {year} {2004})}\BibitemShut {NoStop}%
\bibitem [{\citenamefont {Curtiss}\ and\ \citenamefont
  {Bird}(1980{\natexlab{a}})}]{Curtiss-Bird-1980}%
  \BibitemOpen
  \bibfield  {author} {\bibinfo {author} {\bibfnamefont {C.~F.}\ \bibnamefont
  {Curtiss}}\ and\ \bibinfo {author} {\bibfnamefont {R.~B.}\ \bibnamefont
  {Bird}},\ }\href@noop {} {\bibfield  {journal} {\bibinfo  {journal} {J. Chem.
  Phys.}\ }\textbf {\bibinfo {volume} {74}},\ \bibinfo {pages} {2016} (\bibinfo
  {year} {1980}{\natexlab{a}})}\BibitemShut {NoStop}%
\bibitem [{\citenamefont {Curtiss}\ and\ \citenamefont
  {Bird}(1980{\natexlab{b}})}]{Curtiss-Bird-1980a}%
  \BibitemOpen
  \bibfield  {author} {\bibinfo {author} {\bibfnamefont {C.~F.}\ \bibnamefont
  {Curtiss}}\ and\ \bibinfo {author} {\bibfnamefont {R.~B.}\ \bibnamefont
  {Bird}},\ }\href@noop {} {\bibfield  {journal} {\bibinfo  {journal} {J. Chem.
  Phys.}\ }\textbf {\bibinfo {volume} {74}},\ \bibinfo {pages} {2026} (\bibinfo
  {year} {1980}{\natexlab{b}})}\BibitemShut {NoStop}%
\bibitem [{\citenamefont {Giesekus}(1982)}]{Giesekus-1982}%
  \BibitemOpen
  \bibfield  {author} {\bibinfo {author} {\bibfnamefont {H.}~\bibnamefont
  {Giesekus}},\ }\href@noop {} {\bibfield  {journal} {\bibinfo  {journal} {J.
  Non-Newtonian Fluid Mech.}\ }\textbf {\bibinfo {volume} {11}},\ \bibinfo
  {pages} {69} (\bibinfo {year} {1982})}\BibitemShut {NoStop}%
\bibitem [{\citenamefont {Arsac}\ \emph {et~al.}(1994)\citenamefont {Arsac},
  \citenamefont {Carrot}, \citenamefont {Guillet},\ and\ \citenamefont
  {Revenu}}]{Arsac-Carrot-Guillet-Revenu-1994}%
  \BibitemOpen
  \bibfield  {author} {\bibinfo {author} {\bibfnamefont {A.}~\bibnamefont
  {Arsac}}, \bibinfo {author} {\bibfnamefont {C.}~\bibnamefont {Carrot}},
  \bibinfo {author} {\bibfnamefont {J.}~\bibnamefont {Guillet}}, \ and\
  \bibinfo {author} {\bibfnamefont {P.}~\bibnamefont {Revenu}},\ }\href@noop {}
  {\bibfield  {journal} {\bibinfo  {journal} {J. Non-Newtonian Fluid Mech.}\
  }\textbf {\bibinfo {volume} {55}},\ \bibinfo {pages} {21} (\bibinfo {year}
  {1994})}\BibitemShut {NoStop}%
\bibitem [{\citenamefont {Stephanou}\ \emph {et~al.}(2009)\citenamefont
  {Stephanou}, \citenamefont {Baig},\ and\ \citenamefont
  {Mavrantzas}}]{Stephanou-Baig-Mavrantzas-2009}%
  \BibitemOpen
  \bibfield  {author} {\bibinfo {author} {\bibfnamefont {P.~S.}\ \bibnamefont
  {Stephanou}}, \bibinfo {author} {\bibfnamefont {C.}~\bibnamefont {Baig}}, \
  and\ \bibinfo {author} {\bibfnamefont {V.~G.}\ \bibnamefont {Mavrantzas}},\
  }\href@noop {} {\bibfield  {journal} {\bibinfo  {journal} {J. Rheol.}\
  }\textbf {\bibinfo {volume} {53}},\ \bibinfo {pages} {309} (\bibinfo {year}
  {2009})}\BibitemShut {NoStop}%
\bibitem [{\citenamefont {Beris}\ and\ \citenamefont
  {Edwards}(1994)}]{Beris-Edwards-book}%
  \BibitemOpen
  \bibfield  {author} {\bibinfo {author} {\bibfnamefont {A.~N.}\ \bibnamefont
  {Beris}}\ and\ \bibinfo {author} {\bibfnamefont {B.~J.}\ \bibnamefont
  {Edwards}},\ }\href@noop {} {\emph {\bibinfo {title} {Thermodynamics of
  Flowing Systems}}}\ (\bibinfo  {publisher} {Oxford University Press},\
  \bibinfo {address} {Oxford},\ \bibinfo {year} {1994})\BibitemShut {NoStop}%
\bibitem [{\citenamefont {McPhie}\ \emph {et~al.}(2001)\citenamefont {McPhie},
  \citenamefont {Daivis}, \citenamefont {Snook}, \citenamefont {Ennis},\ and\
  \citenamefont {Evans}}]{McPhie-Daivis-Snook-Ennis-Evans-2001}%
  \BibitemOpen
  \bibfield  {author} {\bibinfo {author} {\bibfnamefont {M.~G.}\ \bibnamefont
  {McPhie}}, \bibinfo {author} {\bibfnamefont {P.~J.}\ \bibnamefont {Daivis}},
  \bibinfo {author} {\bibfnamefont {I.~K.}\ \bibnamefont {Snook}}, \bibinfo
  {author} {\bibfnamefont {J.}~\bibnamefont {Ennis}}, \ and\ \bibinfo {author}
  {\bibfnamefont {D.~J.}\ \bibnamefont {Evans}},\ }\href@noop {} {\bibfield
  {journal} {\bibinfo  {journal} {Physica A}\ }\textbf {\bibinfo {volume}
  {299}},\ \bibinfo {pages} {412} (\bibinfo {year} {2001})}\BibitemShut
  {NoStop}%
\bibitem [{\citenamefont {Baiesi}\ \emph
  {et~al.}(2009{\natexlab{a}})\citenamefont {Baiesi}, \citenamefont {Maes},\
  and\ \citenamefont {Wynants}}]{Baiesi-Maes-Wynants-2009}%
  \BibitemOpen
  \bibfield  {author} {\bibinfo {author} {\bibfnamefont {M.}~\bibnamefont
  {Baiesi}}, \bibinfo {author} {\bibfnamefont {C.}~\bibnamefont {Maes}}, \ and\
  \bibinfo {author} {\bibfnamefont {B.}~\bibnamefont {Wynants}},\ }\href@noop
  {} {\bibfield  {journal} {\bibinfo  {journal} {Phys. Rev. Lett.}\ }\textbf
  {\bibinfo {volume} {103}},\ \bibinfo {pages} {010602} (\bibinfo {year}
  {2009}{\natexlab{a}})}\BibitemShut {NoStop}%
\bibitem [{\citenamefont {Baiesi}\ \emph
  {et~al.}(2009{\natexlab{b}})\citenamefont {Baiesi}, \citenamefont {Maes},\
  and\ \citenamefont {Wynants}}]{Baiesi-Maes-Wynants-2009a}%
  \BibitemOpen
  \bibfield  {author} {\bibinfo {author} {\bibfnamefont {M.}~\bibnamefont
  {Baiesi}}, \bibinfo {author} {\bibfnamefont {C.}~\bibnamefont {Maes}}, \ and\
  \bibinfo {author} {\bibfnamefont {B.}~\bibnamefont {Wynants}},\ }\href@noop
  {} {\bibfield  {journal} {\bibinfo  {journal} {J. Stat. Phys.}\ }\textbf
  {\bibinfo {volume} {137}},\ \bibinfo {pages} {1094} (\bibinfo {year}
  {2009}{\natexlab{b}})}\BibitemShut {NoStop}%
\bibitem [{\citenamefont {Seifert}\ and\ \citenamefont
  {Speck}(2001)}]{Seifert-Speck-2010}%
  \BibitemOpen
  \bibfield  {author} {\bibinfo {author} {\bibfnamefont {U.}~\bibnamefont
  {Seifert}}\ and\ \bibinfo {author} {\bibfnamefont {T.}~\bibnamefont
  {Speck}},\ }\href@noop {} {\bibfield  {journal} {\bibinfo  {journal}
  {Europhys. Lett.}\ }\textbf {\bibinfo {volume} {89}},\ \bibinfo {pages}
  {10007} (\bibinfo {year} {2001})}\BibitemShut {NoStop}%
\bibitem [{\citenamefont {McLeish}(2002)}]{McLeish-2002}%
  \BibitemOpen
  \bibfield  {author} {\bibinfo {author} {\bibfnamefont {T.~C.~B.}\
  \bibnamefont {McLeish}},\ }\href@noop {} {\bibfield  {journal} {\bibinfo
  {journal} {Adv. Phys.}\ }\textbf {\bibinfo {volume} {51}},\ \bibinfo {pages}
  {1379} (\bibinfo {year} {2002})}\BibitemShut {NoStop}%
\bibitem [{\citenamefont {Kr\"{o}ger}(2004)}]{Kroger-2004}%
  \BibitemOpen
  \bibfield  {author} {\bibinfo {author} {\bibfnamefont {M.}~\bibnamefont
  {Kr\"{o}ger}},\ }\href@noop {} {\bibfield  {journal} {\bibinfo  {journal}
  {Phys. Rep.}\ }\textbf {\bibinfo {volume} {390}},\ \bibinfo {pages} {453}
  (\bibinfo {year} {2004})}\BibitemShut {NoStop}%
\bibitem [{\citenamefont {Sarman}\ \emph {et~al.}(1991)\citenamefont {Sarman},
  \citenamefont {Evans},\ and\ \citenamefont
  {Cummings}}]{Sarman-Evans-Cummings-1991}%
  \BibitemOpen
  \bibfield  {author} {\bibinfo {author} {\bibfnamefont {S.}~\bibnamefont
  {Sarman}}, \bibinfo {author} {\bibfnamefont {D.~J.}\ \bibnamefont {Evans}}, \
  and\ \bibinfo {author} {\bibfnamefont {P.~T.}\ \bibnamefont {Cummings}},\
  }\href@noop {} {\bibfield  {journal} {\bibinfo  {journal} {J. Chem. Phys.}\
  }\textbf {\bibinfo {volume} {95}},\ \bibinfo {pages} {8675} (\bibinfo {year}
  {1991})}\BibitemShut {NoStop}%
\bibitem [{\citenamefont {Evans}(1991)}]{Evans-1991}%
  \BibitemOpen
  \bibfield  {author} {\bibinfo {author} {\bibfnamefont {D.~J.}\ \bibnamefont
  {Evans}},\ }\href@noop {} {\bibfield  {journal} {\bibinfo  {journal} {Phys.
  Rev. A}\ }\textbf {\bibinfo {volume} {44}} (\bibinfo {year}
  {1991})}\BibitemShut {NoStop}%
\bibitem [{\citenamefont {Evans}\ \emph {et~al.}(1992)\citenamefont {Evans},
  \citenamefont {Baranyai},\ and\ \citenamefont
  {Sarman}}]{Evans-Baranyai-Sarman-1992}%
  \BibitemOpen
  \bibfield  {author} {\bibinfo {author} {\bibfnamefont {D.~J.}\ \bibnamefont
  {Evans}}, \bibinfo {author} {\bibfnamefont {A.}~\bibnamefont {Baranyai}}, \
  and\ \bibinfo {author} {\bibfnamefont {S.}~\bibnamefont {Sarman}},\
  }\href@noop {} {\bibfield  {journal} {\bibinfo  {journal} {Mol. Phys.}\
  }\textbf {\bibinfo {volume} {76}} (\bibinfo {year} {1992})}\BibitemShut
  {NoStop}%
\bibitem [{\citenamefont {Sarman}\ \emph {et~al.}(1992)\citenamefont {Sarman},
  \citenamefont {Evans},\ and\ \citenamefont
  {Baranyai}}]{Sarman-Evans-Baranyai-1992}%
  \BibitemOpen
  \bibfield  {author} {\bibinfo {author} {\bibfnamefont {S.}~\bibnamefont
  {Sarman}}, \bibinfo {author} {\bibfnamefont {D.~J.}\ \bibnamefont {Evans}}, \
  and\ \bibinfo {author} {\bibfnamefont {A.}~\bibnamefont {Baranyai}},\
  }\href@noop {} {\bibfield  {journal} {\bibinfo  {journal} {Phys. Rev. A}\
  }\textbf {\bibinfo {volume} {46}},\ \bibinfo {pages} {893} (\bibinfo {year}
  {1992})}\BibitemShut {NoStop}%
\bibitem [{\citenamefont {Hunt}\ and\ \citenamefont
  {Todd}(2009)}]{Hunt-Todd-2009}%
  \BibitemOpen
  \bibfield  {author} {\bibinfo {author} {\bibfnamefont {T.~A.}\ \bibnamefont
  {Hunt}}\ and\ \bibinfo {author} {\bibfnamefont {B.~D.}\ \bibnamefont
  {Todd}},\ }\href@noop {} {\bibfield  {journal} {\bibinfo  {journal} {J. Chem.
  Phys.}\ }\textbf {\bibinfo {volume} {131}},\ \bibinfo {pages} {054904}
  (\bibinfo {year} {2009})}\BibitemShut {NoStop}%
\bibitem [{\citenamefont {Kremer}\ and\ \citenamefont
  {Grest}(1990)}]{Kremer-Grest-1990}%
  \BibitemOpen
  \bibfield  {author} {\bibinfo {author} {\bibfnamefont {K.}~\bibnamefont
  {Kremer}}\ and\ \bibinfo {author} {\bibfnamefont {G.~S.}\ \bibnamefont
  {Grest}},\ }\href@noop {} {\bibfield  {journal} {\bibinfo  {journal} {J.
  Chem. Phys.}\ }\textbf {\bibinfo {volume} {92}},\ \bibinfo {pages} {5057}
  (\bibinfo {year} {1990})}\BibitemShut {NoStop}%
\bibitem [{\citenamefont {Baranyai}(2000)}]{Baranyai-2000}%
  \BibitemOpen
  \bibfield  {author} {\bibinfo {author} {\bibfnamefont {A.}~\bibnamefont
  {Baranyai}},\ }\href@noop {} {\bibfield  {journal} {\bibinfo  {journal}
  {Phys. Rev. E}\ }\textbf {\bibinfo {volume} {62}},\ \bibinfo {pages} {5989}
  (\bibinfo {year} {2000})}\BibitemShut {NoStop}%
\bibitem [{\citenamefont {Ilg}(2010)}]{Ilg-2010}%
  \BibitemOpen
  \bibfield  {author} {\bibinfo {author} {\bibfnamefont {P.}~\bibnamefont
  {Ilg}},\ }\href@noop {} {\bibfield  {journal} {\bibinfo  {journal} {J.
  Non-Newtonian Fluid Mech.}\ }\textbf {\bibinfo {volume} {165}},\ \bibinfo
  {pages} {973} (\bibinfo {year} {2010})}\BibitemShut {NoStop}%
\bibitem [{\citenamefont {Ilg}\ and\ \citenamefont
  {Kr\"{o}ger}(2011)}]{Ilg-Kroger-2011}%
  \BibitemOpen
  \bibfield  {author} {\bibinfo {author} {\bibfnamefont {P.}~\bibnamefont
  {Ilg}}\ and\ \bibinfo {author} {\bibfnamefont {M.}~\bibnamefont
  {Kr\"{o}ger}},\ }\href@noop {} {\bibfield  {journal} {\bibinfo  {journal} {J.
  Rheol.}\ }\textbf {\bibinfo {volume} {55}},\ \bibinfo {pages} {69} (\bibinfo
  {year} {2011})}\BibitemShut {NoStop}%
\bibitem [{\citenamefont {Johnson}\ and\ \citenamefont
  {Segalman}(1977)}]{Johnson-Segalman-1977}%
  \BibitemOpen
  \bibfield  {author} {\bibinfo {author} {\bibfnamefont {M.~W.}\ \bibnamefont
  {Johnson}}\ and\ \bibinfo {author} {\bibfnamefont {D.}~\bibnamefont
  {Segalman}},\ }\href@noop {} {\bibfield  {journal} {\bibinfo  {journal} {J.
  Non-Newtonian Fluid Mech.}\ }\textbf {\bibinfo {volume} {2}},\ \bibinfo
  {pages} {255} (\bibinfo {year} {1977})}\BibitemShut {NoStop}%
\bibitem [{\citenamefont {Wu}\ and\ \citenamefont
  {Sch\"{u}mmer}(1990)}]{Wu-Schummer-1990}%
  \BibitemOpen
  \bibfield  {author} {\bibinfo {author} {\bibfnamefont {Q.}~\bibnamefont
  {Wu}}\ and\ \bibinfo {author} {\bibfnamefont {P.}~\bibnamefont
  {Sch\"{u}mmer}},\ }\href@noop {} {\bibfield  {journal} {\bibinfo  {journal}
  {Rheol. Acta}\ }\textbf {\bibinfo {volume} {29}},\ \bibinfo {pages} {23}
  (\bibinfo {year} {1990})}\BibitemShut {NoStop}%
\bibitem [{\citenamefont {Zylka}\ and\ \citenamefont
  {\"{O}ttinger}(1989)}]{Zylka-Ottinger-1989}%
  \BibitemOpen
  \bibfield  {author} {\bibinfo {author} {\bibfnamefont {W.}~\bibnamefont
  {Zylka}}\ and\ \bibinfo {author} {\bibfnamefont {H.~C.}\ \bibnamefont
  {\"{O}ttinger}},\ }\href@noop {} {\bibfield  {journal} {\bibinfo  {journal}
  {J. Chem. Phys.}\ }\textbf {\bibinfo {volume} {90}},\ \bibinfo {pages} {474}
  (\bibinfo {year} {1989})}\BibitemShut {NoStop}%
\bibitem [{\citenamefont {Wedgewood}(1989)}]{Wedgewood-1989}%
  \BibitemOpen
  \bibfield  {author} {\bibinfo {author} {\bibfnamefont {L.~E.}\ \bibnamefont
  {Wedgewood}},\ }\href@noop {} {\bibfield  {journal} {\bibinfo  {journal} {J.
  Non-Newtonian Fluid Mech.}\ }\textbf {\bibinfo {volume} {31}},\ \bibinfo
  {pages} {127} (\bibinfo {year} {1989})}\BibitemShut {NoStop}%
\bibitem [{\citenamefont {Prabhakar}\ and\ \citenamefont
  {Prakash}(2001)}]{Prabhakar-Prakash-2002}%
  \BibitemOpen
  \bibfield  {author} {\bibinfo {author} {\bibfnamefont {R.}~\bibnamefont
  {Prabhakar}}\ and\ \bibinfo {author} {\bibfnamefont {J.~R.}\ \bibnamefont
  {Prakash}},\ }\href@noop {} {\bibfield  {journal} {\bibinfo  {journal} {J.
  Rheol.}\ }\textbf {\bibinfo {volume} {46}},\ \bibinfo {pages} {1191}
  (\bibinfo {year} {2001})}\BibitemShut {NoStop}%
\bibitem [{\citenamefont {Graessley}(1967)}]{Graessley-1967}%
  \BibitemOpen
  \bibfield  {author} {\bibinfo {author} {\bibfnamefont {W.~W.}\ \bibnamefont
  {Graessley}},\ }\href@noop {} {\bibfield  {journal} {\bibinfo  {journal} {J.
  Chem. Phys.}\ }\textbf {\bibinfo {volume} {47}},\ \bibinfo {pages} {1942}
  (\bibinfo {year} {1967})}\BibitemShut {NoStop}%
\bibitem [{\citenamefont {Macdonald}\ and\ \citenamefont
  {Bird}(1966)}]{Macdonald-Bird-1966}%
  \BibitemOpen
  \bibfield  {author} {\bibinfo {author} {\bibfnamefont {I.~F.}\ \bibnamefont
  {Macdonald}}\ and\ \bibinfo {author} {\bibfnamefont {R.~B.}\ \bibnamefont
  {Bird}},\ }\href@noop {} {\bibfield  {journal} {\bibinfo  {journal} {J. Phys.
  Chem.}\ }\textbf {\bibinfo {volume} {70}},\ \bibinfo {pages} {2098} (\bibinfo
  {year} {1966})}\BibitemShut {NoStop}%
\bibitem [{\citenamefont {Tanner}(1968)}]{Tanner-1968}%
  \BibitemOpen
  \bibfield  {author} {\bibinfo {author} {\bibfnamefont {R.~I.}\ \bibnamefont
  {Tanner}},\ }\href@noop {} {\bibfield  {journal} {\bibinfo  {journal} {Trans.
  Soc. Rheol. (J. Rheol.)}\ }\textbf {\bibinfo {volume} {12}},\ \bibinfo
  {pages} {155} (\bibinfo {year} {1968})}\BibitemShut {NoStop}%
\bibitem [{\citenamefont {Yamamoto}(1971)}]{Yamamoto-1971}%
  \BibitemOpen
  \bibfield  {author} {\bibinfo {author} {\bibfnamefont {M.}~\bibnamefont
  {Yamamoto}},\ }\href@noop {} {\bibfield  {journal} {\bibinfo  {journal}
  {Trans. Soc. Rheol. (J. Rheol.)}\ }\textbf {\bibinfo {volume} {15}},\
  \bibinfo {pages} {331} (\bibinfo {year} {1971})}\BibitemShut {NoStop}%
\bibitem [{\citenamefont {Marrucci}(1996)}]{Marrucci-1996}%
  \BibitemOpen
  \bibfield  {author} {\bibinfo {author} {\bibfnamefont {G.}~\bibnamefont
  {Marrucci}},\ }\href@noop {} {\bibfield  {journal} {\bibinfo  {journal} {J.
  Non-Newtonian Fluid Mech.}\ }\textbf {\bibinfo {volume} {62}},\ \bibinfo
  {pages} {279} (\bibinfo {year} {1996})}\BibitemShut {NoStop}%
\bibitem [{\citenamefont {Ianniruberto}\ and\ \citenamefont
  {Marrucci}(1996)}]{Ianniruberto-Marrucci-1996}%
  \BibitemOpen
  \bibfield  {author} {\bibinfo {author} {\bibfnamefont {G.}~\bibnamefont
  {Ianniruberto}}\ and\ \bibinfo {author} {\bibfnamefont {G.}~\bibnamefont
  {Marrucci}},\ }\href@noop {} {\bibfield  {journal} {\bibinfo  {journal} {J.
  Non-Newtonian Fluid Mech.}\ }\textbf {\bibinfo {volume} {65}},\ \bibinfo
  {pages} {241} (\bibinfo {year} {1996})}\BibitemShut {NoStop}%
\bibitem [{\citenamefont {Mead}\ \emph {et~al.}(1998)\citenamefont {Mead},
  \citenamefont {Larson},\ and\ \citenamefont {Doi}}]{Mead-Larson-Doi-1998}%
  \BibitemOpen
  \bibfield  {author} {\bibinfo {author} {\bibfnamefont {D.~W.}\ \bibnamefont
  {Mead}}, \bibinfo {author} {\bibfnamefont {R.~G.}\ \bibnamefont {Larson}}, \
  and\ \bibinfo {author} {\bibfnamefont {M.}~\bibnamefont {Doi}},\ }\href@noop
  {} {\bibfield  {journal} {\bibinfo  {journal} {Macromolecules}\ }\textbf
  {\bibinfo {volume} {31}},\ \bibinfo {pages} {7895} (\bibinfo {year}
  {1998})}\BibitemShut {NoStop}%
\bibitem [{\citenamefont {Ianniruberto}\ and\ \citenamefont
  {Marrucci}(2001)}]{Ianniruberto-Marrucci-2001}%
  \BibitemOpen
  \bibfield  {author} {\bibinfo {author} {\bibfnamefont {G.}~\bibnamefont
  {Ianniruberto}}\ and\ \bibinfo {author} {\bibfnamefont {G.}~\bibnamefont
  {Marrucci}},\ }\href@noop {} {\bibfield  {journal} {\bibinfo  {journal} {J.
  Rheol.}\ }\textbf {\bibinfo {volume} {45}},\ \bibinfo {pages} {1305}
  (\bibinfo {year} {2001})}\BibitemShut {NoStop}%
\bibitem [{\citenamefont {Masubuchi}\ \emph {et~al.}(2001)\citenamefont
  {Masubuchi}, \citenamefont {Takimoto}, \citenamefont {Koyama}, \citenamefont
  {Ianniruberto}, \citenamefont {Greco},\ and\ \citenamefont
  {Marrucci}}]{Masubuchi-Takimoto-Koyama-Ianniruberto-Greco-Marrucci-2001}%
  \BibitemOpen
  \bibfield  {author} {\bibinfo {author} {\bibfnamefont {Y.}~\bibnamefont
  {Masubuchi}}, \bibinfo {author} {\bibfnamefont {J.}~\bibnamefont {Takimoto}},
  \bibinfo {author} {\bibfnamefont {K.}~\bibnamefont {Koyama}}, \bibinfo
  {author} {\bibfnamefont {G.}~\bibnamefont {Ianniruberto}}, \bibinfo {author}
  {\bibfnamefont {F.}~\bibnamefont {Greco}}, \ and\ \bibinfo {author}
  {\bibfnamefont {G.}~\bibnamefont {Marrucci}},\ }\href@noop {} {\bibfield
  {journal} {\bibinfo  {journal} {J. Chem. Phys.}\ }\textbf {\bibinfo {volume}
  {115}},\ \bibinfo {pages} {4387} (\bibinfo {year} {2001})}\BibitemShut
  {NoStop}%
\bibitem [{\citenamefont {Graham}\ \emph {et~al.}(2003)\citenamefont {Graham},
  \citenamefont {Likhtman}, \citenamefont {McLeish},\ and\ \citenamefont
  {Milner}}]{Graham-Likhtman-McLeish-Milner-2003}%
  \BibitemOpen
  \bibfield  {author} {\bibinfo {author} {\bibfnamefont {R.~S.}\ \bibnamefont
  {Graham}}, \bibinfo {author} {\bibfnamefont {A.~E.}\ \bibnamefont
  {Likhtman}}, \bibinfo {author} {\bibfnamefont {T.~C.~B.}\ \bibnamefont
  {McLeish}}, \ and\ \bibinfo {author} {\bibfnamefont {S.~T.}\ \bibnamefont
  {Milner}},\ }\href@noop {} {\bibfield  {journal} {\bibinfo  {journal} {J.
  Rheol.}\ }\textbf {\bibinfo {volume} {47}},\ \bibinfo {pages} {1171}
  (\bibinfo {year} {2003})}\BibitemShut {NoStop}%
\bibitem [{\citenamefont {Inoue}\ and\ \citenamefont
  {Osaki}(1996)}]{Inoue-Osaki-1996}%
  \BibitemOpen
  \bibfield  {author} {\bibinfo {author} {\bibfnamefont {T.}~\bibnamefont
  {Inoue}}\ and\ \bibinfo {author} {\bibfnamefont {K.}~\bibnamefont {Osaki}},\
  }\href@noop {} {\bibfield  {journal} {\bibinfo  {journal} {Macromolecules}\
  }\textbf {\bibinfo {volume} {29}},\ \bibinfo {pages} {1595} (\bibinfo {year}
  {1996})}\BibitemShut {NoStop}%
\bibitem [{\citenamefont {Sridhar}\ \emph {et~al.}(2000)\citenamefont
  {Sridhar}, \citenamefont {Nguyen},\ and\ \citenamefont
  {Fuller}}]{Sridhar-Nguyen-Fuller-2000}%
  \BibitemOpen
  \bibfield  {author} {\bibinfo {author} {\bibfnamefont {T.}~\bibnamefont
  {Sridhar}}, \bibinfo {author} {\bibfnamefont {D.~A.}\ \bibnamefont {Nguyen}},
  \ and\ \bibinfo {author} {\bibfnamefont {G.~G.}\ \bibnamefont {Fuller}},\
  }\href@noop {} {\bibfield  {journal} {\bibinfo  {journal} {J. Non-Newtonian
  Fluid Mech.}\ }\textbf {\bibinfo {volume} {99}},\ \bibinfo {pages} {299}
  (\bibinfo {year} {2000})}\BibitemShut {NoStop}%
\bibitem [{\citenamefont {Luap}\ \emph {et~al.}(2005)\citenamefont {Luap},
  \citenamefont {M\"{u}ller}, \citenamefont {Schweizer},\ and\ \citenamefont
  {Venerus}}]{Luap-Muller-Schweizer-Venerus-2005}%
  \BibitemOpen
  \bibfield  {author} {\bibinfo {author} {\bibfnamefont {C.}~\bibnamefont
  {Luap}}, \bibinfo {author} {\bibfnamefont {C.}~\bibnamefont {M\"{u}ller}},
  \bibinfo {author} {\bibfnamefont {T.}~\bibnamefont {Schweizer}}, \ and\
  \bibinfo {author} {\bibfnamefont {D.~C.}\ \bibnamefont {Venerus}},\
  }\href@noop {} {\bibfield  {journal} {\bibinfo  {journal} {Rheol. Acta}\
  }\textbf {\bibinfo {volume} {45}},\ \bibinfo {pages} {83} (\bibinfo {year}
  {2005})}\BibitemShut {NoStop}%
\bibitem [{\citenamefont {Harada}\ and\ \citenamefont
  {Sasa}(2005)}]{Harada-Sasa-2005}%
  \BibitemOpen
  \bibfield  {author} {\bibinfo {author} {\bibfnamefont {T.}~\bibnamefont
  {Harada}}\ and\ \bibinfo {author} {\bibfnamefont {S.}~\bibnamefont {Sasa}},\
  }\href@noop {} {\bibfield  {journal} {\bibinfo  {journal} {Phys. Rev. Lett.}\
  }\textbf {\bibinfo {volume} {95}},\ \bibinfo {pages} {130602} (\bibinfo
  {year} {2005})}\BibitemShut {NoStop}%
\bibitem [{\citenamefont {Speck}\ and\ \citenamefont
  {Seifert}(2006)}]{Speck-Seifert-2006}%
  \BibitemOpen
  \bibfield  {author} {\bibinfo {author} {\bibfnamefont {T.}~\bibnamefont
  {Speck}}\ and\ \bibinfo {author} {\bibfnamefont {U.}~\bibnamefont
  {Seifert}},\ }\href@noop {} {\bibfield  {journal} {\bibinfo  {journal}
  {Europhys. Lett.}\ }\textbf {\bibinfo {volume} {74}},\ \bibinfo {pages} {391}
  (\bibinfo {year} {2006})}\BibitemShut {NoStop}%
\bibitem [{\citenamefont {Chetrite}\ \emph {et~al.}(2008)\citenamefont
  {Chetrite}, \citenamefont {Falkovich},\ and\ \citenamefont
  {Gaw\c{e}dzki}}]{Chetrite-Falkovich-Gawedzki-2008}%
  \BibitemOpen
  \bibfield  {author} {\bibinfo {author} {\bibfnamefont {R.}~\bibnamefont
  {Chetrite}}, \bibinfo {author} {\bibfnamefont {G.}~\bibnamefont {Falkovich}},
  \ and\ \bibinfo {author} {\bibfnamefont {K.}~\bibnamefont {Gaw\c{e}dzki}},\
  }\href@noop {} {\bibfield  {journal} {\bibinfo  {journal} {J. Stat. Mech.}\
  }\textbf {\bibinfo {volume} {2008}},\ \bibinfo {pages} {P08005} (\bibinfo
  {year} {2008})}\BibitemShut {NoStop}%
\bibitem [{\citenamefont {Chetrite}\ and\ \citenamefont
  {Gaw\c{e}dzki}(2009)}]{Chetrite-Gawedzki-2009}%
  \BibitemOpen
  \bibfield  {author} {\bibinfo {author} {\bibfnamefont {R.}~\bibnamefont
  {Chetrite}}\ and\ \bibinfo {author} {\bibfnamefont {K.}~\bibnamefont
  {Gaw\c{e}dzki}},\ }\href@noop {} {\bibfield  {journal} {\bibinfo  {journal}
  {J. Stat. Phys.}\ }\textbf {\bibinfo {volume} {137}},\ \bibinfo {pages} {890}
  (\bibinfo {year} {2009})}\BibitemShut {NoStop}%
\bibitem [{\citenamefont {Ohta}\ and\ \citenamefont
  {Ohkuma}(2008)}]{Ohta-Ohkuma-2008}%
  \BibitemOpen
  \bibfield  {author} {\bibinfo {author} {\bibfnamefont {T.}~\bibnamefont
  {Ohta}}\ and\ \bibinfo {author} {\bibfnamefont {T.}~\bibnamefont {Ohkuma}},\
  }\href@noop {} {\bibfield  {journal} {\bibinfo  {journal} {J. Phys. Soc.
  Jpn.}\ }\textbf {\bibinfo {volume} {77}},\ \bibinfo {pages} {074004}
  (\bibinfo {year} {2008})}\BibitemShut {NoStop}%
\bibitem [{\citenamefont {Lebowitz}\ and\ \citenamefont
  {Spohn}(1999)}]{Lebowitz-Spohn-1999}%
  \BibitemOpen
  \bibfield  {author} {\bibinfo {author} {\bibfnamefont {J.~L.}\ \bibnamefont
  {Lebowitz}}\ and\ \bibinfo {author} {\bibfnamefont {H.}~\bibnamefont
  {Spohn}},\ }\href@noop {} {\bibfield  {journal} {\bibinfo  {journal} {J.
  Stat. Phys.}\ }\textbf {\bibinfo {volume} {95}},\ \bibinfo {pages} {333}
  (\bibinfo {year} {1999})}\BibitemShut {NoStop}%
\bibitem [{\citenamefont {Qian}(2001)}]{Qian-2001}%
  \BibitemOpen
  \bibfield  {author} {\bibinfo {author} {\bibfnamefont {H.}~\bibnamefont
  {Qian}},\ }\href@noop {} {\bibfield  {journal} {\bibinfo  {journal} {Phys.
  Rev. E}\ }\textbf {\bibinfo {volume} {65}},\ \bibinfo {pages} {016102}
  (\bibinfo {year} {2001})}\BibitemShut {NoStop}%
\bibitem [{\citenamefont {Maes}\ \emph {et~al.}(2008)\citenamefont {Maes},
  \citenamefont {Neto\v{c}n\'{y}},\ and\ \citenamefont
  {Wynants}}]{Maes-Netocny-Wynants-2008}%
  \BibitemOpen
  \bibfield  {author} {\bibinfo {author} {\bibfnamefont {C.}~\bibnamefont
  {Maes}}, \bibinfo {author} {\bibfnamefont {K.}~\bibnamefont
  {Neto\v{c}n\'{y}}}, \ and\ \bibinfo {author} {\bibfnamefont {B.}~\bibnamefont
  {Wynants}},\ }\href@noop {} {\bibfield  {journal} {\bibinfo  {journal}
  {Physica A}\ }\textbf {\bibinfo {volume} {387}},\ \bibinfo {pages} {2675}
  (\bibinfo {year} {2008})}\BibitemShut {NoStop}%
\bibitem [{\citenamefont {Evans}\ and\ \citenamefont
  {Searles}(2002)}]{Evans-Searles-2002}%
  \BibitemOpen
  \bibfield  {author} {\bibinfo {author} {\bibfnamefont {D.~J.}\ \bibnamefont
  {Evans}}\ and\ \bibinfo {author} {\bibfnamefont {D.~J.}\ \bibnamefont
  {Searles}},\ }\href@noop {} {\bibfield  {journal} {\bibinfo  {journal} {Adv.
  Phys.}\ }\textbf {\bibinfo {volume} {51}},\ \bibinfo {pages} {1529} (\bibinfo
  {year} {2002})}\BibitemShut {NoStop}%
\bibitem [{\citenamefont {Blickle}\ \emph {et~al.}(2007)\citenamefont
  {Blickle}, \citenamefont {Speck}, \citenamefont {Lutz}, \citenamefont
  {Siefert},\ and\ \citenamefont
  {Bechinger}}]{Blickle-Speck-Lutz-Seifert-Bechinger-2007}%
  \BibitemOpen
  \bibfield  {author} {\bibinfo {author} {\bibfnamefont {V.}~\bibnamefont
  {Blickle}}, \bibinfo {author} {\bibfnamefont {T.}~\bibnamefont {Speck}},
  \bibinfo {author} {\bibfnamefont {C.}~\bibnamefont {Lutz}}, \bibinfo {author}
  {\bibfnamefont {U.}~\bibnamefont {Siefert}}, \ and\ \bibinfo {author}
  {\bibfnamefont {C.}~\bibnamefont {Bechinger}},\ }\href@noop {} {\bibfield
  {journal} {\bibinfo  {journal} {Phys. Rev. Lett.}\ }\textbf {\bibinfo
  {volume} {98}},\ \bibinfo {pages} {210601} (\bibinfo {year}
  {2007})}\BibitemShut {NoStop}%
\bibitem [{\citenamefont {Gomez-Solano}\ \emph {et~al.}(2009)\citenamefont
  {Gomez-Solano}, \citenamefont {Petrosyan}, \citenamefont {Ciliberto},
  \citenamefont {Chetrite},\ and\ \citenamefont
  {Gaw\c{e}dzki}}]{GomezSolano-Petrosyan-Ciliberto-Chetrite-Gawedzki-2009}%
  \BibitemOpen
  \bibfield  {author} {\bibinfo {author} {\bibfnamefont {J.~R.}\ \bibnamefont
  {Gomez-Solano}}, \bibinfo {author} {\bibfnamefont {A.}~\bibnamefont
  {Petrosyan}}, \bibinfo {author} {\bibfnamefont {S.}~\bibnamefont
  {Ciliberto}}, \bibinfo {author} {\bibfnamefont {R.}~\bibnamefont {Chetrite}},
  \ and\ \bibinfo {author} {\bibfnamefont {K.}~\bibnamefont {Gaw\c{e}dzki}},\
  }\href@noop {} {\bibfield  {journal} {\bibinfo  {journal} {Phys. Rev. Lett.}\
  }\textbf {\bibinfo {volume} {103}},\ \bibinfo {pages} {040601} (\bibinfo
  {year} {2009})}\BibitemShut {NoStop}%
\bibitem [{\citenamefont {Marrucci}(1985)}]{Marrucci-1985}%
  \BibitemOpen
  \bibfield  {author} {\bibinfo {author} {\bibfnamefont {G.}~\bibnamefont
  {Marrucci}},\ }\href@noop {} {\bibfield  {journal} {\bibinfo  {journal} {J.
  Polym. Sci. B: Polym. Phys.}\ }\textbf {\bibinfo {volume} {23}},\ \bibinfo
  {pages} {159} (\bibinfo {year} {1985})}\BibitemShut {NoStop}%
\bibitem [{\citenamefont {Watanabe}\ \emph
  {et~al.}(2003{\natexlab{b}})\citenamefont {Watanabe}, \citenamefont {Ishida},
  \citenamefont {Matsumiya},\ and\ \citenamefont
  {Inoue}}]{Watanabe-Ishida-Matsumiya-Inoue-2004}%
  \BibitemOpen
  \bibfield  {author} {\bibinfo {author} {\bibfnamefont {H.}~\bibnamefont
  {Watanabe}}, \bibinfo {author} {\bibfnamefont {S.}~\bibnamefont {Ishida}},
  \bibinfo {author} {\bibfnamefont {Y.}~\bibnamefont {Matsumiya}}, \ and\
  \bibinfo {author} {\bibfnamefont {T.}~\bibnamefont {Inoue}},\ }\href@noop {}
  {\bibfield  {journal} {\bibinfo  {journal} {Macromolecules}\ }\textbf
  {\bibinfo {volume} {37}},\ \bibinfo {pages} {6619} (\bibinfo {year}
  {2003}{\natexlab{b}})}\BibitemShut {NoStop}%
\bibitem [{\citenamefont {Watanabe}\ \emph {et~al.}(2006)\citenamefont
  {Watanabe}, \citenamefont {Sawada},\ and\ \citenamefont
  {Matsumiya}}]{Watanabe-Sawada-Matsumiya-2006}%
  \BibitemOpen
  \bibfield  {author} {\bibinfo {author} {\bibfnamefont {H.}~\bibnamefont
  {Watanabe}}, \bibinfo {author} {\bibfnamefont {T.}~\bibnamefont {Sawada}}, \
  and\ \bibinfo {author} {\bibfnamefont {Y.}~\bibnamefont {Matsumiya}},\
  }\href@noop {} {\bibfield  {journal} {\bibinfo  {journal} {Macromolecules}\
  }\textbf {\bibinfo {volume} {39}},\ \bibinfo {pages} {2553} (\bibinfo {year}
  {2006})}\BibitemShut {NoStop}%
\bibitem [{\citenamefont {Watanabe}(2008)}]{Watanabe-2008}%
  \BibitemOpen
  \bibfield  {author} {\bibinfo {author} {\bibfnamefont {H.}~\bibnamefont
  {Watanabe}},\ }\href@noop {} {\bibfield  {journal} {\bibinfo  {journal}
  {Prog. Theor. Phys. Suppl.}\ }\textbf {\bibinfo {volume} {175}},\ \bibinfo
  {pages} {17} (\bibinfo {year} {2008})}\BibitemShut {NoStop}%
\bibitem [{\citenamefont {Shanbhag}\ \emph {et~al.}(2001)\citenamefont
  {Shanbhag}, \citenamefont {Larson}, \citenamefont {Takimoto},\ and\
  \citenamefont {Doi}}]{Shanbhag-Larson-Takimoto-Doi-2001}%
  \BibitemOpen
  \bibfield  {author} {\bibinfo {author} {\bibfnamefont {S.}~\bibnamefont
  {Shanbhag}}, \bibinfo {author} {\bibfnamefont {R.~G.}\ \bibnamefont
  {Larson}}, \bibinfo {author} {\bibfnamefont {J.}~\bibnamefont {Takimoto}}, \
  and\ \bibinfo {author} {\bibfnamefont {M.}~\bibnamefont {Doi}},\ }\href@noop
  {} {\bibfield  {journal} {\bibinfo  {journal} {Phys. Rev. Lett.}\ }\textbf
  {\bibinfo {volume} {87}},\ \bibinfo {pages} {195502} (\bibinfo {year}
  {2001})}\BibitemShut {NoStop}%
\bibitem [{\citenamefont {Phan-Thien}\ \emph {et~al.}(1984)\citenamefont
  {Phan-Thien}, \citenamefont {Manero},\ and\ \citenamefont
  {Leal}}]{PhanThien-Manero-Leal-1984}%
  \BibitemOpen
  \bibfield  {author} {\bibinfo {author} {\bibfnamefont {N.}~\bibnamefont
  {Phan-Thien}}, \bibinfo {author} {\bibfnamefont {O.}~\bibnamefont {Manero}},
  \ and\ \bibinfo {author} {\bibfnamefont {L.~G.}\ \bibnamefont {Leal}},\
  }\href@noop {} {\bibfield  {journal} {\bibinfo  {journal} {Rheol. Acta}\
  }\textbf {\bibinfo {volume} {23}} (\bibinfo {year} {1984})}\BibitemShut
  {NoStop}%
\bibitem [{\citenamefont {Baxandall}(1987)}]{Baxandall-1987}%
  \BibitemOpen
  \bibfield  {author} {\bibinfo {author} {\bibfnamefont {L.~G.}\ \bibnamefont
  {Baxandall}},\ }\href@noop {} {\bibfield  {journal} {\bibinfo  {journal} {J.
  Chem. Phys.}\ }\textbf {\bibinfo {volume} {87}},\ \bibinfo {pages} {2297}
  (\bibinfo {year} {1987})}\BibitemShut {NoStop}%
\bibitem [{\citenamefont {Biller}\ and\ \citenamefont
  {Petruccione}(1988)}]{Biller-Petruccione-1988}%
  \BibitemOpen
  \bibfield  {author} {\bibinfo {author} {\bibfnamefont {P.}~\bibnamefont
  {Biller}}\ and\ \bibinfo {author} {\bibfnamefont {F.}~\bibnamefont
  {Petruccione}},\ }\href@noop {} {\bibfield  {journal} {\bibinfo  {journal}
  {J. Chem. Phys.}\ }\textbf {\bibinfo {volume} {89}},\ \bibinfo {pages} {2412}
  (\bibinfo {year} {1988})}\BibitemShut {NoStop}%
\bibitem [{\citenamefont {Likhtman}(2005)}]{Likhtman-2005}%
  \BibitemOpen
  \bibfield  {author} {\bibinfo {author} {\bibfnamefont {A.~E.}\ \bibnamefont
  {Likhtman}},\ }\href@noop {} {\bibfield  {journal} {\bibinfo  {journal}
  {Macromolecules}\ }\textbf {\bibinfo {volume} {38}},\ \bibinfo {pages} {6128}
  (\bibinfo {year} {2005})}\BibitemShut {NoStop}%
\bibitem [{\citenamefont {Gordon}\ and\ \citenamefont
  {Schowalter}(1972)}]{Gordon-Showalter-1972}%
  \BibitemOpen
  \bibfield  {author} {\bibinfo {author} {\bibfnamefont {R.~J.}\ \bibnamefont
  {Gordon}}\ and\ \bibinfo {author} {\bibfnamefont {W.~R.}\ \bibnamefont
  {Schowalter}},\ }\href@noop {} {\bibfield  {journal} {\bibinfo  {journal}
  {Trans. Soc. Rheol. (J. Rheol.)}\ }\textbf {\bibinfo {volume} {16}},\
  \bibinfo {pages} {79} (\bibinfo {year} {1972})}\BibitemShut {NoStop}%
\bibitem [{\citenamefont {Gardiner}(2004)}]{Gardiner-book}%
  \BibitemOpen
  \bibfield  {author} {\bibinfo {author} {\bibfnamefont {C.~W.}\ \bibnamefont
  {Gardiner}},\ }\href@noop {} {\emph {\bibinfo {title} {Handbook of Stochastic
  Methods}}},\ \bibinfo {edition} {3rd}\ ed.\ (\bibinfo  {publisher}
  {Springer},\ \bibinfo {address} {Berlin},\ \bibinfo {year}
  {2004})\BibitemShut {NoStop}%
\bibitem [{\citenamefont {Kleinert}(2004)}]{Kleinert-book}%
  \BibitemOpen
  \bibfield  {author} {\bibinfo {author} {\bibfnamefont {H.}~\bibnamefont
  {Kleinert}},\ }\href@noop {} {\emph {\bibinfo {title} {Path Integrals in
  Quantum Mechanics, Statistics, Polymer Physics, and Finantial Markets}}},\
  \bibinfo {edition} {3rd}\ ed.\ (\bibinfo  {publisher} {World Scientific},\
  \bibinfo {address} {Singapore},\ \bibinfo {year} {2004})\BibitemShut
  {NoStop}%
\end{thebibliography}%

\clearpage

\section*{Figure Captions}

\hspace{-\parindent}%
FIG. \ref{schematic_image_polymer_under_shear}. Schematic image of a
 polymer chain under shear. The solid curve and the solid arrow
 represents the polymer
 chain and its end-to-end vector of the chain, respectively.
 The dotted arrows represent the flow field given by eq \eqref{velocity_gradient_tensor_simple_shear}. $\bm{R}$ and $\dot{\gamma}$
 are the end-to-end vector and the shear rate, respectively.

\

\hspace{-\parindent}%
FIG. \ref{eta_and_psi1_power_law}. The shear viscosity
 $\eta(\dot{\gamma})$ and the first normal stress difference coefficient
 $\Psi_{1}(\dot{\gamma})$ calculated by the power law type model
 (eq \eqref{dimensionless_mobility_factor_power_law}). The exponent is
 set to $\alpha = 9 / 11$.

\

\hspace{-\parindent}%
FIG. \ref{parallel_moduli_power_law} The parallel storage and loss
 moduli $G'_{\parallel}(\dot{\gamma},\omega)$ and $G''_{\parallel}(\dot{\gamma},\omega)$, calculated by the power law type mobility model
 (eq \eqref{dimensionless_mobility_factor_power_law}). The exponent is
 set to $\alpha = 9 / 11$.

\

\hspace{-\parindent}%
FIG. \ref{entropy_production_rate_power_law} The entropy production
 rate $\Sigma(\dot{\gamma})$ calculated by the power law type model
 (eq \eqref{dimensionless_mobility_factor_power_law}). The exponent and
 the characteristic crossover time are
 set to $\alpha = 9 / 11$ and $\tau_{c} = \tau_{0}$, respectively.

\

\hspace{-\parindent}%
FIG. \ref{rheo_dielectric_weak_coupling} The rheo-dielectric
 response functions for the weak kinetic coupling model. The
 effective coupling constant is varied as $a (\tau_{0} \dot{\gamma})^{2}
 = 0, 0.1, 0.2$. The case of $a (\tau_{0} \dot{\gamma})^{2} = 0$
 corresponds to the equilibrium dielectric response.

\clearpage

\section*{Figures}

\end{document}